\documentclass[journal]{IEEEtran}
\usepackage[utf8]{inputenc}

\usepackage{amsmath,amsfonts}
\usepackage{pifont}
\usepackage[caption=false,font=normalsize,labelfont=sf,textfont=sf]{subfig}
\usepackage{url}
\usepackage{graphicx}
\usepackage{cite}

\usepackage{graphicx}				
\usepackage{amssymb}
\usepackage{bm}
\usepackage{siunitx}
\usepackage{ifthen}
\usepackage[dvipsnames]{xcolor}   
\usepackage{hyperref}
\hypersetup{
	colorlinks=true, 
	linktoc=all,     
	linkcolor=blue,  
	citecolor=black,
}

\newcommand{\myVec}[1]{\bm{#1}}

\newcommand{\vecex}{\myVec{e}_x}

\newcommand{\veck}{\myVec{k}}
\newcommand{\vecU}{\myVec{U}}

\def\del{\partial}

\newcommand{\DSV}{\tilde{c}}
\newcommand{\vecDSV}{\tilde{\myVec{c}}}

\newcommand{\myMean}[1]{\left\langle #1 \right\rangle}

\DeclareMathOperator{\sinc}{sinc}
\newcommand{\Angg}{ \theta}
\newcommand{\myvartheta}{ \Angg}
\newcommand{\Ang}{ \theta_0}
\newcommand{\FT}[1]{\mathcal{F}\{#1\}}
\newcommand{\DAng}{\Delta \Angg}

\newcommand{\dcdk}{\delta c_{\delta k}}
\newcommand{\dcdw}{\delta c_{\delta \omega}}

\def\myFigWidthSingles{0.49\textwidth}

\newcommand{\caseName}[1]{%
\ifthenelse{#1=1}{case-$\Ang$-$\DAng$}{%
\ifthenelse{#1=2}{case-$\gamma$}{%
\ifthenelse{#1=3}{case-$L$-$T$}{%
\ifthenelse{#1=4}{case-NSP-LS}{%
\ifthenelse{#1=5}{case-$U$}{%
"name XX" 
}}}}}}
\newcommand{\caseAng}[1]{%
\ifthenelse{#1=1}{\ang{60}, \ang{90}}{%
\ifthenelse{#1=2}{\ang{60}}{%
\ifthenelse{#1=3}{\ang{90}}{%
\ifthenelse{#1=4}{\ang{75}}{%
\ifthenelse{#1=5}{\ang{90}}{%
XX 
}}}}}}
\newcommand{\caseDang}[1]{%
\ifthenelse{#1=1}{\ang{15}...\ang{60}}{%
\ifthenelse{#1=2}{\ang{15}...\ang{60}}{%
\ifthenelse{#1=3}{\ang{60}}{%
\ifthenelse{#1=4}{\ang{60}}{%
\ifthenelse{#1=5}{\ang{60}}{%
XX 
}}}}}}
\newcommand{\caseGamma}[1]{
\ifthenelse{#1=1}{3.3}{%
\ifthenelse{#1=2}{1...7}{%
\ifthenelse{#1=3}{3.3}{%
\ifthenelse{#1=4}{3.3}{%
\ifthenelse{#1=5}{3.3}{%
XX 
}}}}}}
\newcommand{\caseL}[1]{%
\ifthenelse{#1=1}{10}{%
\ifthenelse{#1=2}{10}{%
\ifthenelse{#1=3}{5, 10, 20}{%
\ifthenelse{#1=4}{10}{%
\ifthenelse{#1=5}{10}{%
XX 
}}}}}}
\newcommand{\caseT}[1]{%
\ifthenelse{#1=1}{20}{%
\ifthenelse{#1=2}{20}{%
\ifthenelse{#1=3}{10, 20, 80}{%
\ifthenelse{#1=4}{20}{%
\ifthenelse{#1=5}{20}{%
XX 
}}}}}}
\newcommand{\caseU}[1]{%
\ifthenelse{#1=1}{0}{%
\ifthenelse{#1=2}{0}{%
\ifthenelse{#1=3}{0}{%
\ifthenelse{#1=4}{0}{%
\ifthenelse{#1=5}{0...0.15}{%
XX 
}}}}}}
\newcommand{\caseBatch}[1]{%
\ifthenelse{#1=1}{b20220903 yeolde}{%
\ifthenelse{#1=2}{b20230206 gamma}{%
\ifthenelse{#1=3}{b20230114 LnT}{%
\ifthenelse{#1=4}{b20220903 yeolde}{%
\ifthenelse{#1=5}{b20230203 constCurr}{%
XX 
}}}}}}
\newcommand{\caseFig}[1]{%
\ifthenelse{#1=1}{\ref{fig:spec_ex}}{%
\ifthenelse{#1=2}{\ref{fig:peakedness}}{%
\ifthenelse{#1=3}{\ref{fig:video_duration}}{%
\ifthenelse{#1=4}{\ref{fig:ex_NSP_vs_LS}}{%
\ifthenelse{#1=5}{\ref{fig:ConstCurr}, \ref{fig:ConstCurr_rot30}}{%
XX 
}}}}}}
\newcommand{\caseSec}[1]{%
\ifthenelse{#1=1}{\ref{sec:DSV_example},\ref{sec:angwidth}}{%
\ifthenelse{#1=2}{\ref{sec:DSV_gamma}}{%
\ifthenelse{#1=3}{\ref{sec:size}}{%
\ifthenelse{#1=4}{\ref{sec:LSvsNSP}}{%
\ifthenelse{#1=5}{\ref{sec:DSV_U}}{%
XX 
}}}}}}

\DeclareRobustCommand{\caseNameRobust}[1] { \caseName{#1} }

\begin{document}
	
	\title{
		Biases from spectral leakage in remote sensing of near-surface currents}
	\author{Stefan Weichert, Benjamin K. Smeltzer, Simen \AA. Ellingsen 
		\thanks{The work of S.~\AA.\ Ellingsen was supported in part by the European Union (ERC, WaTurSheD, project 101045299) and the Research Council of Norway (project 325114). Views and opinions expressed are however those of the authors only and do not necessarily reflect those of the European Union or the European Research Council. Neither the European Union nor the granting authority can be held responsible for them.}
		\thanks{S.~Weichert and S.~\AA.\ Ellingsen are with the Department of Energy and Process Engineering, Norwegian University of Science and Technology, 7491 Trondheim, Norway.} 
		\thanks{B.~K.\ Smeltzer is with SINTEF Ocean, Marinteknisk senter, 7052 Trondheim, Norway}
	}
	
	\maketitle
	
	\begin{abstract}
		Remotely measuring subsurface water currents from imagery of the wave field has become a much-used technique. We study the biases and errors in such measurements due to spectral leakage, and suggest mitigating procedures.
		Deviations between peak values in the three-dimensional wave spectrum and the known dispersion relation in quiescent water are extracted and interpreted as current-induced Doppler shifts, from which the sub-surface current is inferred. 
		The use of discrete Fourier transforms, however, introduces  spectral leakage between nearby frequency bins. 
		Analysing synthetically generated wave data adhering to realistic input spectra we show that although no current is in fact present, spurious currents can be ``measured'' which can amount to a significant fraction of the phase speed at the spectral peak. 
		We analyse the effects of data tapering, method of Doppler shift extraction, limited wavenumber and frequency resolution, peakedness and angular width of the input spectrum, and average misalignment between waves and Doppler shift velocity direction.
		The narrower the input wave spectrum in frequency and/or direction, the greater the biases become. The use of a window function reduces the severity in nearly all cases, yet mitigates the effects of limited resolution more effectively in space than in time.      
		When a current is present the absolute biases remain essentially unchanged, when waves and currents are roughly aligned 
        tapering much alleviates the concomitant biases%
        , whereas in the case of a cross-current, biases remain significant even for tapered data.		
	\end{abstract}


	\section{Introduction}
	
	The prospect of measuring currents near the sea surface remotely from above is a highly attractive one. Measuring depth-varying currents \emph{in situ} by penetrating the surface requires the use of e.g.\ buoys, ships, gliders or fixed instruments, all relatively expensive and able to measure a single point or trajectory at a time, and often struggle to capture currents in the top few metres. In comparison, remote measurement from above can be performed with inexpensive equipment mounted on airborne 
	platforms
	able to cover larger areas in a short time (see, e.g., \cite{Smeltzer21}). 
	
	By far the most common source of wave data
 for this purpose
 has been measurements using HF or X-band radar \cite{Stewart74,Ha79,Young85,Fernandez96,Gurgel00,Teague01,Gangeskar02,Ardhuin09,Huang12,Hessner14,Lund15,Huang16,Lund18,Gangeskar18, stole-hentschel23}, primarily mounted on ships. Only observation of the wave phase variation in time and space is required, however, meaning other methods are equally applicable; the use of infrared \cite{Dugan96} and polarimetric \cite{Zappa12,Laxague17} imaging has been demonstrated, as has regular optical measurements (video) with cameras mounted on quadcopter drones \cite{Stresser17}, and aircraft \cite{Dugan01,Dugan03,Yurovskaya18,Lenain23}.
	
	A current varying with depth will affect the wave phase velocity differently for different wavelengths, resulting in a measurable effective Doppler shift which depends on wavenumber \cite{Stewart74} and appears as a shift in the waves' spectral dispersion curve from that observed in quiescent water. 
	Ever more advanced methods have been developed in recent years for inferring the depth profile of the sub-surface currents from such measured Doppler shifts \cite{Lund15,Campana17,Smeltzer19}, yet the task of obtaining these shifts from an observation of the spatiotemporal wave field, by reconstructing the current-modified dispersion surface in the frequency-wavenumber spectrum, is itself a nontrivial task. Typically, the spectrum is divided into wavenumber magnitude bins, whereby the Doppler shifts are found for each bin separately. Perhaps the most frequently employed are least-squares-based methods (e.g.\ \cite{Young85,Gangeskar02,Senet01}), whereas the alternative Normalized Scalar Product (NSP) method is also in regular use (e.g.\ \cite{Huang12,Serafino10,Huang16}). We compare these methods herein finding NSP to be unequivocally favourable. A further method which we shall not consider here is the so-called Polar Current Shell method (e.g.\ \cite{Huang16}), recently adapted for this purpose \cite{Smeltzer19}, which has similar performance as NSP, but favourable in some circumstances.

	One should note that the same questions we seek to answer here, also apply to bathymetry retrieval from the measured wave spectra (see e.g.\ \cite{lund20} and references therein).
	The extraction of the water depth from the measured spectrum requires spectral intensity at low wavenumbers, which can be strongly influenced by spectral leakage, as we will show. 
	Since a finite water depth enters the dispersion relation in a direction-independent and multiplicative way in contrast to the additive and anisotropic term due to a current, it is not obvious how the issues discussed herein affect the bathymetry retrieval quantitatively, a question of potential importance which requires further investigation.

	\subsection{Outline}
	
	In this work we consider the effects of a strong spectral energy peak and the corresponding effect it has on the extraction of Doppler shifts for wavenumbers in the vicinity of the peak. The energy peak results in spectral leakage to adjacent wavevectors in the wave spectrum and may cause errors in the extracted Doppler shifts --- as perhaps the clearest example we show that in realistic seastates, spectral leakage can cause a significant spurious current to be ``measured'' when none is in fact present. 
	
	The phenomenon of spectral leakage is briefly reviewed in section \ref{sec:leakage}, and 
	we discuss how it can be mitigated by data windowing. We then go on to describe a numerical experiment where mock wave data is generated and analysed with the methods in standard use for remote-sensing of currents from wave dispersion, in section \ref{sec:methods}. Results are reported studying  how the spurious current ``measurements'' depend on the method of Doppler shift extraction (section \ref{sec:doppler}), the angular width and peakedness of a directional 
 wave spectrum of the Joint North Sea Wave Project type (JONSWAP) \cite{Hasselmann73}
 (sections \ref{sec:angwidth} and \ref{sec:peakedness}), the spatial and temporal resolution (sections \ref{sec:size} and \ref{sec:duration}), as well as how the situation changes when a uniform background current is present, in section \ref{sec:current}. 
	We finally summarize and give a brief overview of ways whereby the detrimental effect of spectral leakage can be reduced in practical applications, in section \ref{sec:conclusions}.

	\section{Background}\label{sec:background}
	
	When remote sensing of sub-surface currents from wave dispersion is performed, the input spectrum is an observation of the motion of the water surface resolved in time and space. The method is based on linear wave theory, so  only the phase of the waves is required, not the amplitude. The data is typically a monotonic function of the sea surface elevation or its derivative, as a function of position $(x,y)$ and time $t$.
	A three dimensional Fourier transform is then applied in space and time to obtain a spectral signal as a function of wave numbers $\veck=(k_x, k_y)$ and frequency $\omega$. We assume the spatial area and duration of the observation are $L\times L$ and $T$, respectively, so that the resolution in wave-number and frequency are, respectively, $\delta k=2\pi/L$ and $\delta \omega=2\pi/T$; we will refer to these as a pixel or bin in wavenumber and frequency, respectively. (Note that after we introduce nondimensional units in section \ref{sec:dimensions} $\delta k$ and $\delta \omega$ take the forms $1/L$ and $1/T$, respectively.)  We will assume infinitely deep water for simplicity herein.
	
	\subsection{Theory}\label{sec:theory}
	
	In a wave spectrum, the spectral signal is concentrated near the dispersion relation $\omega = \omega_\text{DR}(\veck)$. The methods for sensing the sub-surface current now extract a measured function $\omega(\veck)$. If a current with moderately strong depth-dependence $\vecU(z) = (U(z),V(z))$ is present, the dispersion relation for a wave with wave vector $\veck$ is well approximated as  $\omega
	=\omega_\text{DR}(\veck;\vecDSV)$ with the dispersion function
	\begin{equation}\label{eq:SnJ}
		\omega_\text{DR}(\veck;\vecDSV) 
		= \omega_0(k)+\veck\cdot \vecDSV
	\end{equation}
	where $k=|\veck|$, $\vecDSV$ is the Doppler shift velocity (DSV) due to the presence of a sub-surface current and the dispersion relation in deep, quiescent water is 
	\begin{equation}
		\omega_0(k) = \sqrt{gk}.
	\end{equation}
	
	Inversion methods to infer depth-dependent velocity profiles from the measured spectrum are based on the approximation \cite{Stewart74,Ell17}
	\begin{equation}\label{eq:SJ}
		\vecDSV=
		2k\int_{-\infty}^0 \vecU(z) e^{2kz} dz.
	\end{equation}
	If a current is present which is uniform in depth, the resulting Doppler shift will be independent of $\veck$, while conversely, a Doppler shift which varies with $\veck$ implies the presence of a current which varies as a function of $z$. We shall see that biases in the measured Doppler shift $\vecDSV$ due to spectral leakage typically vary significantly with $\veck$, and hence the spurious currents which are ``measured'' will have a nontrivial depth dependence. We do not pursue this question in detail.
	
	The presence of the current thus introduces an observable Doppler shift corresponding to the addition of a phase speed $\vecDSV(k)$ to the phase velocity.
	Remote sensing of the depth-varying current $\vecU(z)$ is then possible by measuring $\vecDSV$ and inverting equation \eqref{eq:SJ} using one of several methods available as reviewed in \cite{Smeltzer21}.
	Methods for extracting $\vecDSV$ from a measured spectrum are reviewed and compared in section \ref{sec:doppler}. 
	
	Following Smeltzer et al.\ \cite{Smeltzer19} we define the instructive quantities  $\dcdw$ and $\dcdk$ as
	\begin{equation}\label{eq:dcdwk}
		\dcdw=\frac{\delta\omega}{k}; ~~~  \dcdk=\frac{\del\omega}{\del k} \frac{\delta k}{k}, 
	\end{equation}
	which estimate the change in the predicted phase velocity $c$ due to moving the dispersion surface \eqref{eq:SnJ} by $\delta k$ along the wavenumber axis in the spectrum, or by $\delta\omega$ along the frequency axis. These are thus approximate measures of the uncertainty in velocity measurement introduced by limited wavenumber and frequency resolution, respectively. Further discussion may be found in section 4.2.1 of \cite{Smeltzer19}.

	\subsection{Spectral leakage and windowing}\label{sec:leakage}
	\label{sec:window}
	
	Assume a continuous signal (in space or time) $\tilde{P}_0(t)$, the ``true'' signal, is measured during a finite period of duration $T$. A sharp cut-off at the beginning and end of the measurement is equivalent to multiplying $\tilde{P}_0$ by a discontinuous top-hat function $w_{\text{box}}(t)$ which is $1$ within a time interval of length $T$ and zero outside this ``window''. 
	Multiplication with such a window is equivalent to a convolution of the 
	spectrum ${P}_0 = \FT{\tilde{P}_0}$ with a $\sinc$ function in frequency space (e.g.\ \cite{Lyon09}), 
	\begin{equation}
		P_{\text{measured}}(f) = P_0(f)*T\sinc(fT),
	\end{equation}
	where $\sinc(a)=\sin(\pi a)/\pi a$,
	which is the Fourier transform of the top hat function. The result is a blurring of the spectrum:
	``true" spectral components will appear as spectral intensity not only in the frequency bins closest to the actual frequency of said component, but also on neighbouring ones that are more than one bin away, with an intensity decreasing with distance. 
	One can alleviate this by replacing the ``box" window with a window of choice, simply by multiplying the acquired data $\tilde{P}(t) = \tilde{P}_0 w_\text{box}(t)$ with said window 
	function $w(t)$. If the chosen $w(t)$ also vanishes outside the measurement interval, the window replaces the top-hat. The choice of the optimal window function much depends on the situation and data at hand.

	\begin{figure}[ht]
		\centering
		\includegraphics[width = .48\textwidth]{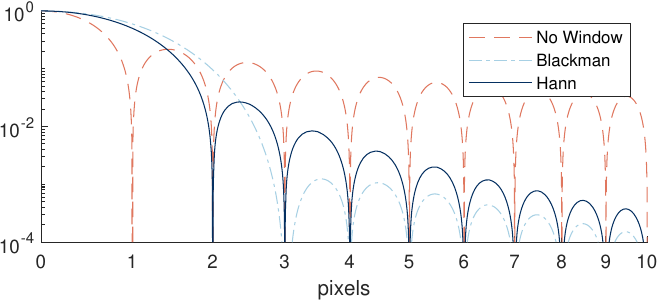}
		\caption{
			Fourier transform $\FT{w}(f)$ of a selection of common window functions $w(t)$. A frequency pixel is $\delta \omega = 2\pi/T$, normalized so that $\FT{w}(0)=1$. This illustrates how long-range spectral leakage can be mitigated by tapering the measured signal. The drawback of increased short-range leakage (1-2 pixels) is often negligible.
		}
		\label{fig:leakage_lobes}
	\end{figure}

	The Fourier transforms $\FT{w}(f)$ of some common window functions are shown in figure \ref{fig:leakage_lobes} (See e.g.\ \cite{Nuttall1981} for a more comprehensive comparison).
	When taking a discrete Fourier transform (DFT) of $w(t)$ every real frequency component $f$ will generate spectral intensity in frequency bins of width $\delta f$ centered at $n\delta f$, $n\in Z$, with  intensity $\FT{w}(n\delta f-f)$. For example, for the \emph{Blackman window} shown in figure \ref{fig:leakage_lobes}, a frequency coinciding with a DFT frequency, i.e. $f=m\delta f$, one obtains non-zero spectral intensity only for $n-m=0,1,2$, leading to no long-range leakage whatsoever. Only the central lobe is sampled in frequency space. For any real spectrum most frequencies will not coincide with a DFT frequency, and so the side-lobes give rise to long-range leakage.
	Now, the central lobe of the \emph{No Window} (or top hat) case is only one pixel wide, but the leakage is very long-range, i.e.\ spectral intensity decays slowly and stays well above  $1\%$ for more than $20$ pixels. 
	In contrast to this, the \emph{Blackman} window decays rapidly to $10^{-3}$, at the cost of increasing spectral leakage into 
	the two nearest bins, 1 and 2. 
	The best choice of the windowing function is usually a compromise between suppressing long-range leakage and blurring the spectrum (short-range leakage). 
	In this work, a Hann window, defined as
	\begin{equation}\label{eq:hann}
		w_{\text{H}}(t)  = 0.5- 0.5\,\cos\left(2\pi\,t/T\right)  ,
	\end{equation}
	is used, as it suppresses the long-range spectral leakage to less than 1\% and leaks significantly into frequencies less than two pixels away. In appendix \ref{app:window_comparison} an example of extracted DSVs with different windowing functions is given to illustrate how the choice affects the results.
	
	The data in this work is surface elevation $\zeta(x,y,t)$ measured in three dimensions
	$x$, $y$ and time $t$, which we pre-multiply by a 3D Hann window constructed as 
	$   w(x,y,t)=w_{\text{H}}(x)w_{\text{H}}(y)w_{\text{H}}(t)$
	prior to subjecting it to a discrete 3-dimensional fast-Fourier transform (3DFFT)
 (it is understood that $L$ replaces $T$ in equation \eqref{eq:hann} when the argument is $x$ or $y$%
 . 
	The  new signal $\zeta(x,y,t) w(x,y,t)$ goes smoothly to zero at the edges of the domain of observation, in our case the square area with sides $L$ and time duration $T$.

	\section{Methods} \label{sec:methods}
	
	We proceed by producing synthetic surface elevation data $\zeta(x,y,t)$ by superposing random linear plane waves of  wave number $k$  and direction $\Angg$ from chosen spectra with varying properties, as detailed in section \ref{sec:spectrum}. Each wave is given a uniformly distributed random initial phase, the frequency $\omega$  is found from equation \eqref{eq:SnJ}, whereupon the waves are propagated in time.

	These wave ``observations'' are of course idealized, since different methods for obtaining the actual surface elevation $\zeta(x,y,t)$ or true spectrum from field data each come with their individual challenges and limitations, an ongoing field of research in its own right. Taking such practical challenges into account is beyond the scope of this work, and we use the ideal data to isolate the effects of spectra leakage in the data analysis.

	We mostly consider the case of quiescent water, i.e., there is no background current and any Doppler shifts ``measured'' from the spectrum are spurious and purely a consequence of spectral leakage. We also consider the case where a constant background current $\myVec{U}_0$ is present, in section \ref{sec:current}. Biases now manifest as deviations of the observed DSVs from the correct value.

	\subsection{Nondimensional quantities}\label{sec:dimensions}
	Dimensional quantities will be denoted by a superscript asterisk, all other quantities are non-dimensionalized. The  reference length-scale and time-scale are defined based on a characteristic wavenumber $k_0^*$ and its corresponding angular frequency in quiescent water $\omega_0^*=\sqrt{gk_0^*}$. Thus, e.g. $T=T^*\omega_0^*/2\pi$, $L = L^*k_0^*/2\pi$, $k = k^*/k^*_0$, $\omega = \omega^*/\omega_0^*$, $U = U^*\sqrt{k_0^*/g}$. Unless specified otherwise, $\omega_0^*(k_0^*)$ is taken to be the location of the peak of the energy spectrum%
 , i.e., the spectral peak is at $\omega=1$ by definition%
 , hence the peak in wavenumber space is close to to $k=1$.

	\subsection{Wave spectrum}\label{sec:spectrum}
	
\begin{figure}[ht]
		\centering
		\includegraphics[width=0.49\textwidth]{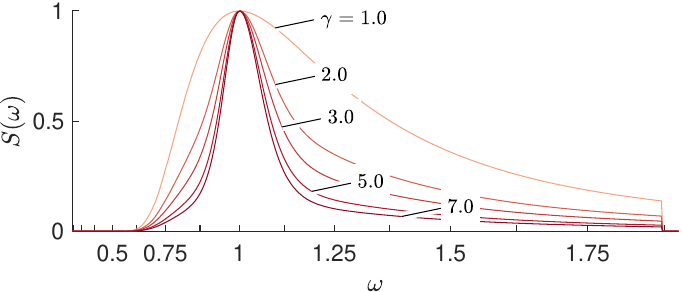}\\
		\includegraphics[width=0.49\textwidth]{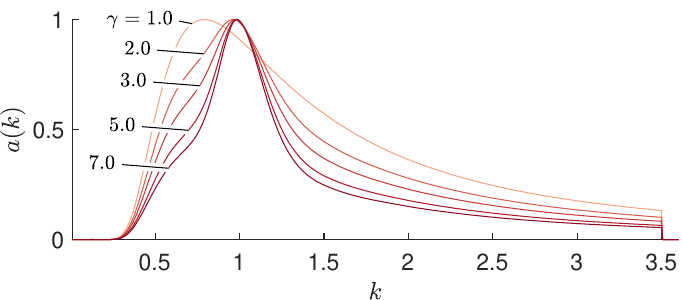}
		\caption{Top: Energy spectrum $S(\omega)$ (equation \ref{eq:JONSWAP}) for a few values of the peakedness parameter $\gamma$. The $\omega$ axis is scaled quadratically to match the range of the bottom graph, as
  $k^* = \omega^{*2}/g$, i.e., 
  $k=\omega^2$. Bottom: amplitudes $a(k,\Angg=\Ang)$ (equation \ref{eq:amplitudes}) of superposed wave components. The spectra are normalized with respect to their peak value.
		}
		\label{fig:spectrum_examples}
	\end{figure}

	We generate wave fields from commonly used realistic model spectra with varying directional broadness, assuming the form 
	\begin{equation}
		\hat{S}(\omega,\theta) = S(\omega)f(\theta)
	\end{equation}
	where $\theta$ is the angle between $\veck$ and the $x$ axis, $\cos\theta=\veck\cdot\vecex/k$. We use the JONSWAP spectrum \cite{Hasselmann73}
	\begin{equation}\label{eq:JONSWAP}
		S(\omega) = \tilde{N} \omega^{-5}
		\exp{\left[-\frac{5}{4}\omega^{-4}\right]\gamma^{r(\omega)}}
	\end{equation}
	with
	\begin{equation}
		r(\omega) =  
		\exp{\left[-\frac{1}{2}\left(\frac{\omega-1}{\sigma}\right)^2\right]}.
	\end{equation}
	For our purposes the value of $\tilde{N}$
	is not of importance, we set $\tilde{N}=1$.
	The parameter $\sigma$ is
	\begin{equation}\label{eq:JONSWAP_params}
		\sigma =
		\begin{cases}
			0.07,&{\text{if}}\ \omega\le \omega_p,\\
			{0.09,}&{\text{if}}\ {\omega > \omega_p}
		\end{cases}
	\end{equation}
	and the peakedness parameter $\gamma$ is varied (see section \ref{sec:DSV_gamma}). The energy spectrum for a selection of peakedness values $\gamma$ is depicted in figure \ref{fig:spectrum_examples}.
	The angular distribution is taken as a cosine-square with a full width (distance of first roots) of  $\DAng$, i.e. 
	\begin{equation}\label{eq:angularspread}
		f(\Angg)= \left\{ \begin{matrix}			\cos^2\left(\pi\frac{\Angg-\Ang}{\DAng}\right) &,\, |\Angg-\Ang|\le \DAng/2  \\			0 &\, \text{else}		\end{matrix} \right. \, \quad \,	
	\end{equation}
	
	We use this spectrum to prescribe the amplitudes $a(\veck)$ of the superposed wave components in accordance with \cite{SocquetJuglard2005} as
	\begin{equation}\label{eq:amplitudes}
		a(\veck) = \sqrt{2\,k^{-3/2} \,\hat{S}(\omega(k),\Angg)\delta k}
	\end{equation}
	on an evenly spaced grid in $k_x$-$k_y$, with spacing $\delta k_x=\delta k_y=\delta k$.
	Additionally, we set $a(\veck$)=0 for $k/k_p>3.5$.

	\subsection{Normalized scalar product} \label{sec:doppler}
	\label{sec:NSP}
	
	Two methods are in common use for extracting Doppler shifts from the measured wave spectrum; least-squares (LS) methods (e.g.\ \cite{Young85,Gangeskar02,Senet01}) and the Normalized Scalar Product (NSP) method (e.g.\ \cite{Huang12,Serafino10,Huang16}). 
	While there exists a wide range of extensions and sophistications to LS methods, we focus our attention on the NSP method. However, since two of the most famous works on spectrum based current extraction (\cite{Young85,Gangeskar02}) use a simple form of the LS method we also give a brief comparison of their performance in Appendix 
 \ref{sec:LSvsNSP}.
	
	We employ the DSV extraction method as implemented by Smeltzer et al.\ \cite{Smeltzer19}; see \cite{Huang12,Huang16,Serafino10} for details on the NSP method more generally. 
	
	The starting point in either method is a measured, spatio-temporally resolved free-surface $\eta(x,y,t)$ and its power spectrum obtained via a discrete Fourier transform, $P(\veck, \omega) = |FFT[\eta(x,y,t)]|^2$. 
	For each wavenumber $k_i$ in a list, the spectral intensity on a cylindrical surface  with radius $k_i$ centered around the $\omega$-axis is defined
	\begin{equation}
		F_i(\veck, \omega) = 
		\sqrt{P(k_i \cos\myvartheta,k_i  \sin\myvartheta, \omega)}.
		\label{eq:maskedSpec}
	\end{equation}
	where $\myvartheta$ is the azimuth angle in the $k_x$-$k_y$-plane.  
	The implementation of the algorithms used are formulated in Cartesian coordinates; however, it is illustrative to use cylindrical coordinates for the following conceptual considerations.
	More details of the implementation are given in  appendix \ref{app:numdetails}. 
	
	To find the effective DSV $\vecDSV_i=\vecDSV(k_i)$, first, a characteristic function $G$ is defined that contains the components of $\vecDSV_i$ as free parameters:
	\begin{equation} \label{eq:DSVkernel}
		G_i(\myvartheta,\omega;\vecDSV_i) = G^+_i(\myvartheta,\omega;\vecDSV_i)+G^-_i(\myvartheta,\omega;\vecDSV_i)
	\end{equation}
	where
	\begin{equation} 
		G^\pm_i(\myvartheta,\omega;\vecDSV_i)= \exp\left[-2\left(    \frac{\omega\pm\omega_\text{DR}(\myvartheta;k_i,\vecDSV_i)}{a}     \right)^2 \right].
	\end{equation}
	The normalized scalar product $N_i$ of the vectors $F_i$ and $G_i$ is now maximized for each value of $i$ by varying the two components of $\vecDSV_i$; it is calculated as
	\begin{equation}\label{eq:normSp}
		N_i(\vecDSV)= \frac{\myMean{G_i F_i}}{\myMean{G_i}\myMean{F_i}}
	\end{equation}
	where $\myMean{...}$ refers to an integral over all $\myvartheta$ and $\omega$.
	In other words, $G_i$ can be thought of as a cosine in $\myvartheta$  (see figure \ref{fig:cylinderIllustration}) with offset $\omega_0(k_i)$ amplitude $k_i \DSV$ and phase shift defined by the direction of $\vecDSV$; its overlap with the measured intensity on the cylinder surface is maximized to find the best DSV $\vecDSV(k_i)$.
	This optimization step is performed using the Nelder-Mead simplex method \cite{Lagarias1998}.
	
	\section{The normalized scalar product method and spectral leakage}\label{sec:NSP_and_leakage}
	
\begin{figure}[ht]
		\centering
		\includegraphics[width=0.48\textwidth]{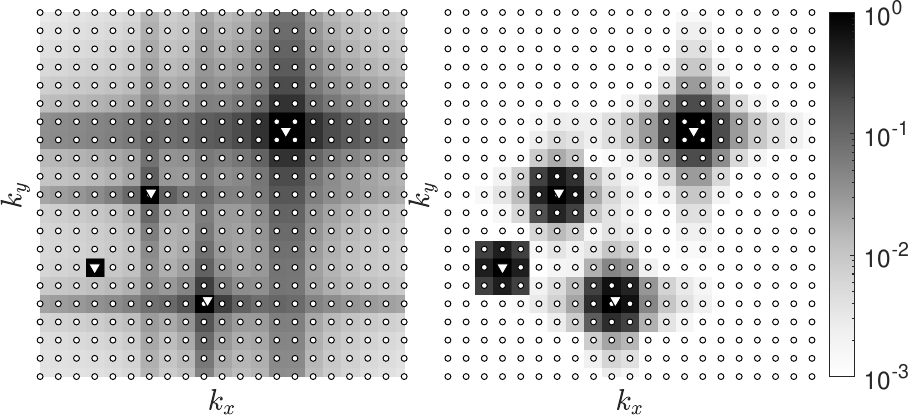}
		\caption{An illustration of spectral leakage in two dimensions. Left: Logarithmic 2D FFT spectrum of a superposition of four sine-waves
  (i.e., the spectrum is the sum of four delta functions in the $\boldsymbol{k}$ plane)%
  , with frequencies denoted by triangles. The circles denote the DFT frequencies, i.e. frequencies natural to the domain.
			From left to right, the leftmost spectral peak coincides with a DFT frequency, while the rightmost peak lies in the middle of DFT frequencies.
			Right: same as left, bit with a 2D Hann window applied before doing the FFT.
			Note how the Hann window removes the background and increases consistency between spectral peak, while broadening by roughly one bin in all directions.
		}
		\label{fig:leakage_illustration2D}
	\end{figure}

	For multidimensional data, spectral leakage is most prominent in the directions parallel to the coordinate axes. Consider for example a 2D signal $\eta(x,y)$ on a rectangular domain and its Fourier transform $\tilde{\eta}(k_x,k_y)=\FT{\eta}$.  Since both $x$ and 
 $y$ 
 are within a finite range, the effective window $w(x,y)$ is a product of top-hat windows in the $x$ and $y$ directions, respectively, i.e. $w = w_1(x)w_2(y)$. 
	The Fourier transform of such a product is the product of their respective Fourier transforms $\tilde{w}(k_x,k_y) = \FT{w_1(x)}\FT{w_2(y)}=\tilde{w}_1(k_x)\tilde{w}_2(k_y) $. 
	Now, since the leakage for a top-hat window falls off as $1/k$, the product $\tilde{w}(k_x,k_y)$  is smallest for a given $k$ when $|k_x|=|k_y|$ and largest when $k_x=0$ or $k_y=0$.
	An illustration of this can be seen in  figure \ref{fig:leakage_illustration2D}.
	The extension to 3 or more dimensions is straightforward.
	
	This (mostly) axes-parallel leakage is helpful in understanding how leakage affects the DSV extraction using NSP. 
	
\begin{figure}[ht]
		\centering
		\includegraphics[width=0.24\textwidth,trim=15 20 20 50 mm, clip=true]{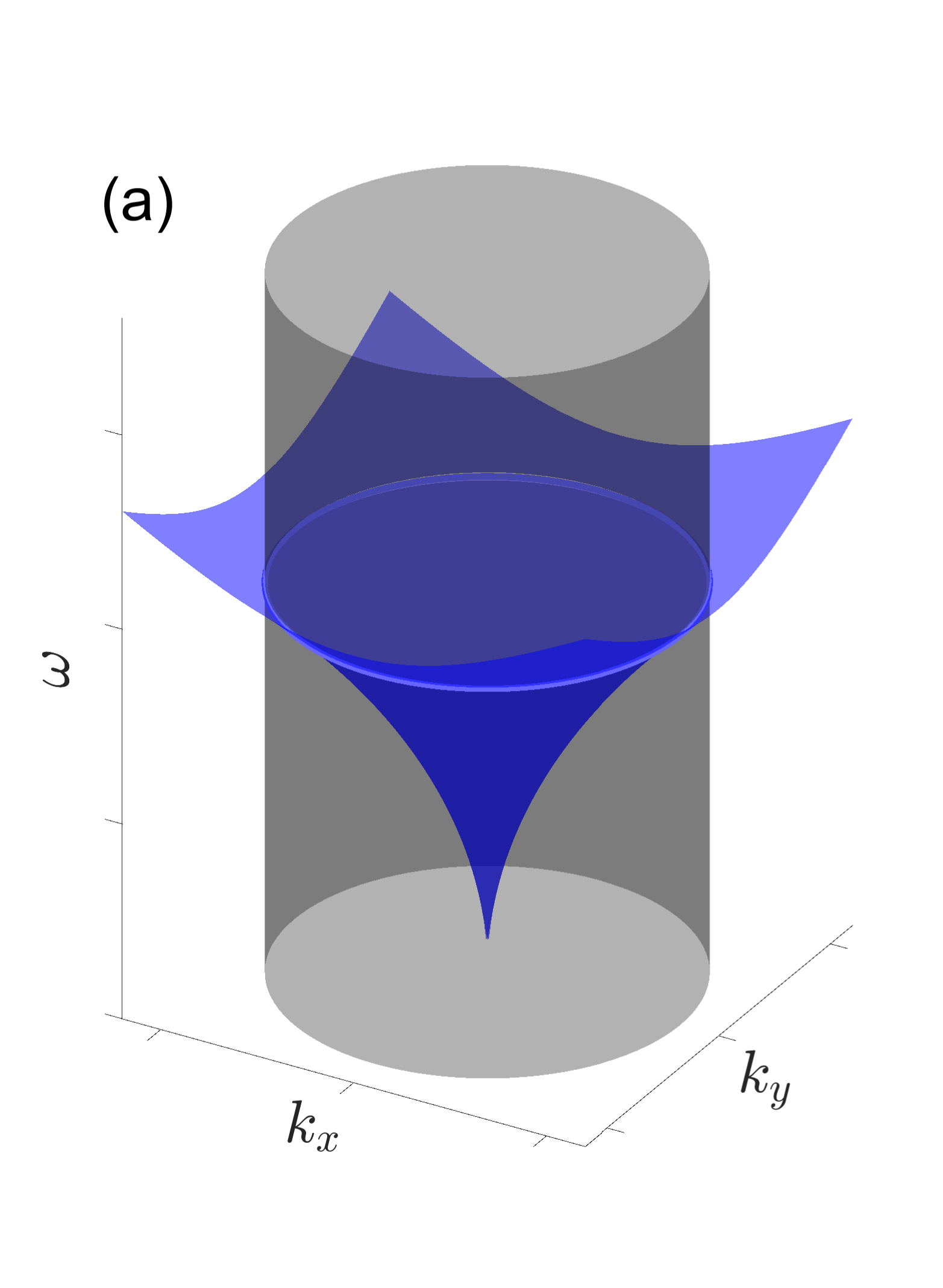}
		\includegraphics[width=0.24\textwidth,trim=15 20 20 50 mm, clip=true]{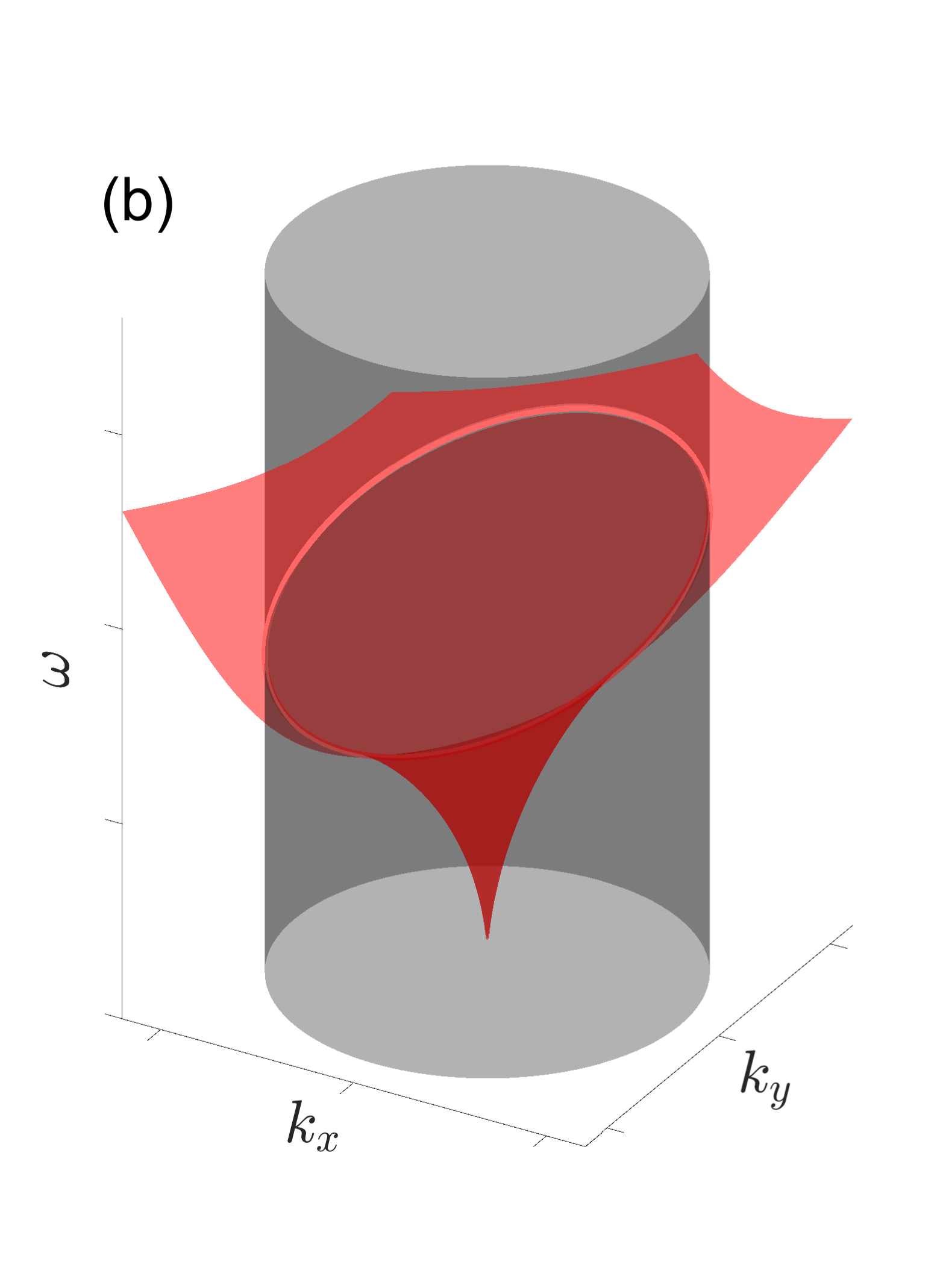}\\
		\includegraphics[width=0.43\textwidth]{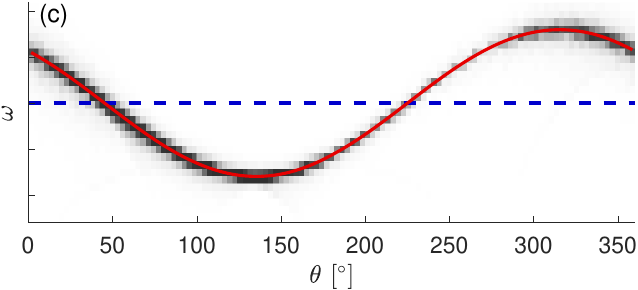}    
		\caption{Illustration of the dispersion relation $\omega =\omega_\text{DR}(\veck;\vecDSV)$ in connection within the Nonlinear Scalar Product method. The top graphs show the dispersion relation in absence of a current, $\vecDSV=0$ (a) and with a current, $\vecDSV=\myVec{U}_0$ (b). The intersections with a cylinder with radius $k_i$ is highlighted. 
			In graph (c) the spectral intensity is indicated in greyscale, visible near the dispersion curve on the cylinder surface after unrolling it,  together with the lines of intersection from the top figures: blue, dashed  corresponds to quiescent water; red, solid is the best fit from the NSP method resulting in a non-vanishing Doppler shift velocity.}
		\label{fig:cylinderIllustration}
	\end{figure}

	A fundamental step in the NSP method is to pick out measured spectral intensity on a cylinder surface with radius $k_i$. 
	One can think of the effect on the NSP as the spectral intensity projecting itself in the principal directions onto the cylinder, with decreasing intensity the further the surface is from the originating spectral intensity.
	Now, as the NSP method essentially fits a function of the form $\omega_0(k_i)+A_f\cos(\Angg-\Angg_f)$, with free parameters $A_f$ and $\Angg_f$, to the spectral intensity on the cylinder surface defined by $k_i$, as illustrated in figure \ref{fig:cylinderIllustration},   spurious intensity at $\omega\neq \omega_0(k_i)$ leads to non-zero values for $A_f=\veck\cdot\vecDSV$, implying a background current even in the absence of one.
	In this work we consider spectra with a single peak at a given wave vector $\veck=(k_x,k_y)$.
	
	We now consider the effect of spectral leakage for two distinct cases, one (1) with the mean propagation direction $\Ang$ being parallel to the $k_y$-axis
 ($\theta_0=90^\circ$)%
 , and one (2) with $\Ang$ between \ang{0} and \ang{90}.

	\subsection{Leakage for \texorpdfstring{$\Ang=\ang{90}$\,}{ang90}}
	Consider the extraction of the DSV for a wavenumber $k<1$, i.e. below the 
 $k$-value at the
 peak of the spectrum
 which is close to $k=1$ (see Section \ref{sec:dimensions}%
 . The spectral leakage for the cylinder with radius $k$ 
 (see Fig.~\ref{fig:cylinderIllustration})
 will be dominated by short-range spectral leakage of wavenumbers close to $k$ and long-range leakage from the spectral peak. 
	the short-range spectral leakage for the most part will 
 merely
 blur the spectral intensity on the cylinder, while the long-range leakage from the spectral peak leads to a new, possibly dominating, spectral intensity spot on the cylinder, at the same angle, but a higher frequency. In other words right above the real spectral intensity.
	Depending on the ratio of intensities, distance, and width of the characteristic function $G$, the NSP algorithm may find the correct frequency, the frequency of the projected peak or a value in between these two to give the best fit. Therefore the returned DSV will be parallel to the spectrum propagation direction $\Ang$.
	Similarly, if $k>1$ is considered, the projection from the spectral peak will be below the real spectral intensity, giving DSV antiparallel to the spectrum propagation direction $\Ang$.

	\subsection{Leakage for \texorpdfstring{$0<\Ang<\ang{90}$\,}{ang0to90}}
	
	For  $k<1$, it is now possible for spectral leakage from the peak to have no axes-parallel projection onto the cylinder surface, thus not affecting the extracted DSV for this wavenumber.
	Short-range spectral leakage and leakage from lower wavenumbers now dominate, making the prediction of DSV direction difficult.
	For $k>1$, one can still get an intuition for the effect of leakage.
	As the peak of the spectrum lies in the first quadrant of the $k_x-k_y$-plane, the shortest distance to a cylinder with radius $k$ is to the quarter of the cylinder also lying in the first quadrant, and we can focus our consideration on that.
	The leakage is now being projected onto the cylinder at an angle relative to the mean propagation direction $\Ang$.
	Therefore, as the  real spectral intensity now dominates for $\myvartheta = \Ang$, the frequency for wavenumbers along $\Ang$ is found correctly.
	The perpendicular component of the DSV, however is determined mostly by the position and relative magnitude of the spurious intensity appearing at angles $\myvartheta\neq\Ang$. 
	in quiescent water, for example, one would obtain DSVs with directions $\myvartheta \pm \ang{90}$.
	
	Note that these considerations hold true for very narrow spectra, but are to be understood as tendencies for spectra with considerable spectral width $\DAng$, because, for any angle $\myvartheta$, real spectral intensity will mostly dominate over spectral leakage, if present.

	\section{Parameter Choice and Overview}\label{sec:parameters}
	
	For an overview and easier referencing table \ref{tab:case_collection} contains the list of test cases presented in this paper.
	Apart from the parameters stated therein, the resolution of the input spectrum $\delta_{in}k$ and spatial resolution $\delta x=\delta y$ as well as temporal resolution $\delta t$ had to be set. The spatial resolution was chosen such that waves of the highest wavenumbers, $k_{max}=4$, would be well resolved, $\delta x = 1/28$. Similarly, time resolution was set to $\delta t=1/14$. The resolution of the input spectrum was set to $\delta_{in} k\approx 0.341/L$, with L being the domain size. This was deemed small enough to mimic a continuous spectrum. 
	
	The domain size and video duration were chosen to be 
 similar to typical parameters found in airborne measurements. For instance, videos used in \cite{Lenain2023}
typically have a field of view of $L^*= \SIrange{128}{512}{m}$, durations of $T^*=\SIrange{20}{40}{s}$ and wavenumbers with usable spectral intensity from \SIrange[range-phrase= { }to{ } ]{0.2}{2}{\radian \per \metre} (see figure 1 in \cite{Lenain2023}).
	Taking a reference wavenumber of $k^*_0=\SI{0.4}{\radian \per \metre}$, this gives a parameter range of $T\approx 5-10$, $L\approx 8-32$. 
	
	Equation \ref{eq:DSVkernel} contains one free parameter, $a$, that determines the width of the characteristic function. We set $a=4\,\delta\omega$, with $\,\delta\omega=1/T$ being the frequency resolution (see appendix \ref{app:numdetails} for details).
	
	The results shown in figures 
 \ref{fig:spec_ex}-\ref{fig:ConstCurr_rot30}
  display statistics of the extracted DSVs in terms of the average $\myMean{|\vecDSV|}$ and the corresponding standard deviation $\sigma_{\DSV}$ (times two for illustrative purposes), calculated from 100 realizations. These quantities are useful as they represent a mean bias and fluctuation that one has to expect from a single measurement. Additionally, where possible, the implied velocity resolutions 
	$\dcdw$ and $\dcdk$ defined in equation \eqref{eq:dcdwk}
	are shown for reference.
	
	\begin{table*}[ht]
		
\newcommand{\caseMakeLine}[1]{
\caseName{#1} &\caseAng{#1} & \caseDang{#1} &  \caseGamma{#1} & \caseL{#1} &\caseT{#1}   &\caseU{#1} & \caseFig{#1} & \caseSec{#1}\\
}

\centering
\begin{tabular}{|l|c|c|c|c|c|c|l|l|}\hline
\rule{0pt}{2ex}   
     & $\Ang$  &   $\DAng$        & $\gamma$ &$L$    & $T$  & $U$ & figure & section\\\hline
      \caseMakeLine{4}
      \caseMakeLine{1}
      \caseMakeLine{2}
      \caseMakeLine{3}
      \caseMakeLine{5}
     \hline
\end{tabular}\\
		\vspace{0.1cm}
		\caption{Overview of parameter combinations. $\Ang$: mean propagation direction of input wavenumber spectrum, $\DAng$: angular spread of input wavenumber spectrum (see equation \ref{eq:angularspread}), $\gamma$: peakedness parameter (see equation \ref{eq:JONSWAP}), $L$: domain spatial domain size, $T$: video duration, $U$: background current. All parameters are non-dimensional, see section \ref{sec:dimensions}}
		\label{tab:case_collection}
	\end{table*}

	\section{Results}
	
	In this section we consider the effects of 
	wave-spectral properties, resolution and data tapering on the spurious Doppler shifts ``measured'' when no current is present, as well as the effect of a uniform current being present. 
	An overview of the parameter combinations used to obtain the following results is given  in section \ref{sec:parameters}.

	\subsection{Illustration of the effect of windowing (Hann window)}
	\label{sec:DSV_example}
	
\begin{figure*}[ht]
		\centering
		\includegraphics[width = \myFigWidthSingles]{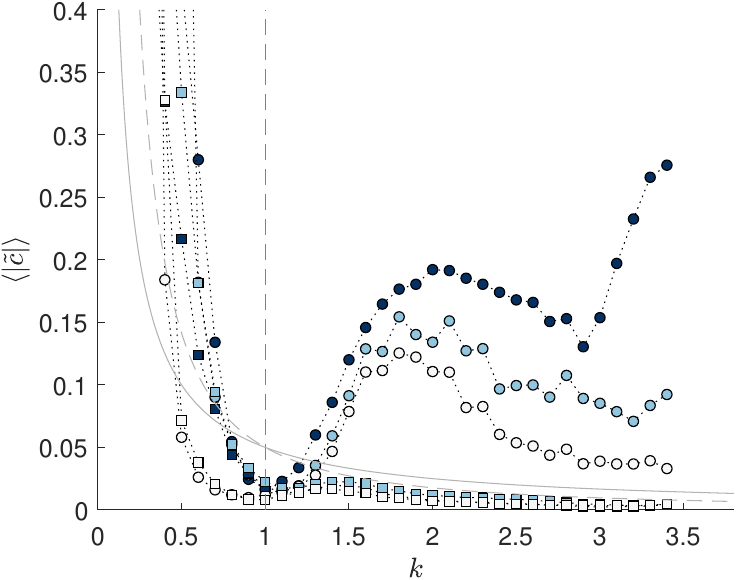}
		\includegraphics[width = \myFigWidthSingles]{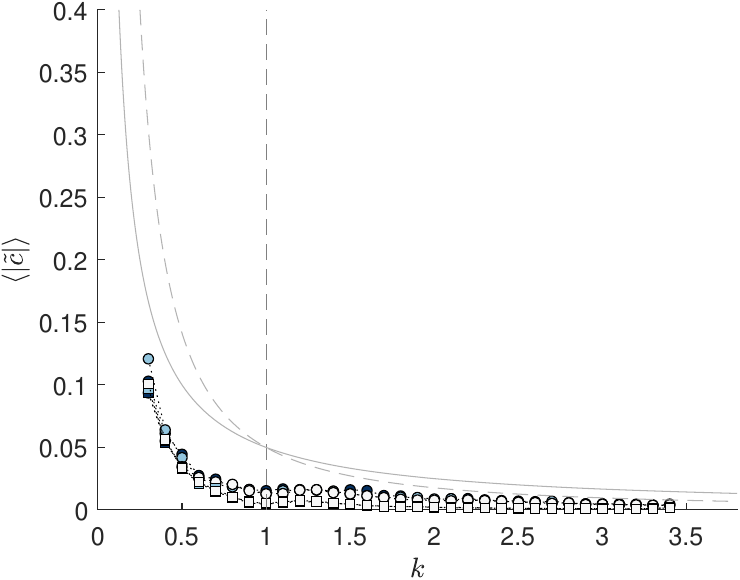} \\
		\includegraphics[width = \myFigWidthSingles]{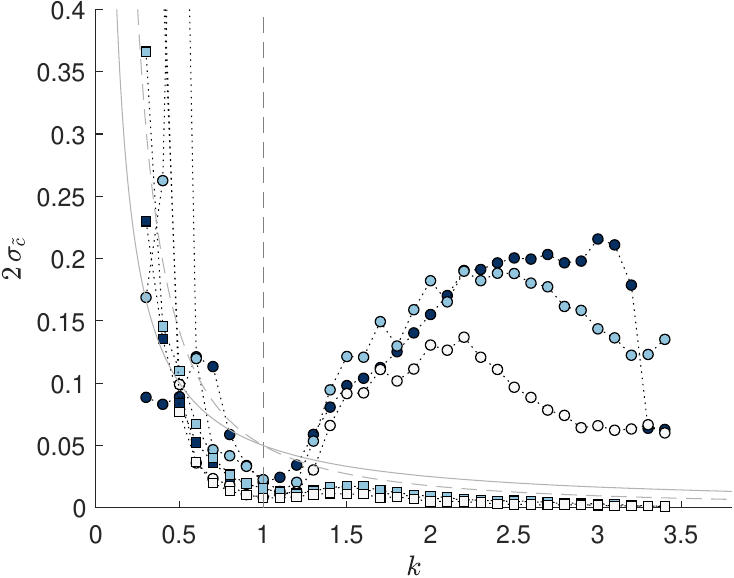}
		\includegraphics[width = \myFigWidthSingles]{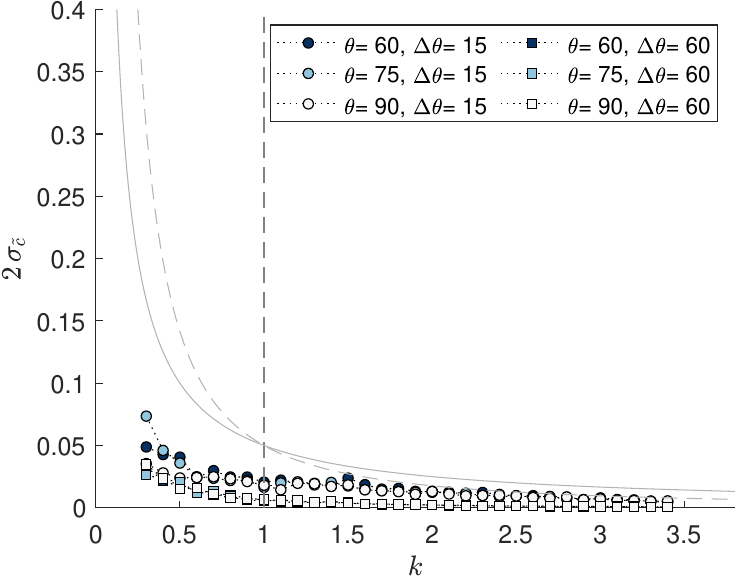}
		\caption{ Doppler shift velocities (DSV) $\vecDSV$
  in units of the peak phase velocity
  for the parameter set of \caseNameRobust{1} (see table \ref{tab:case_collection}) extracted using the NSP method. The DSVs were extracted for 100 realizations with (right) and without (left) applying a Hann window beforehand. The top and bottom graphs show the biases in DSVs in terms of the average (top) and standard deviation (bottom) of the absolute value $\DSV=|\vecDSV|$. The solid and dashed line correspond to the velocity resolutions implied by  frequency and wavenumber resolution, respectively.}
		\label{fig:spec_ex} \label{fig:Ang_Dang}
	\end{figure*}

 The mitigating effect of windowing/tapering is well illustrated when considering
 the influence of the angular spread $\DAng$ of the wavenumber spectrum; simulations with variation both in $\DAng$ and the mean propagation direction $\Ang$ were performed (see table \ref{tab:case_collection}, \caseName{1}, for all parameters). 
	A selection of the results are shown in figure \ref{fig:spec_ex} (upper row). 
	A more detailed discussion of the influence of $\DAng$ and $\Ang$ is given in section \ref{sec:DSV_Dang} below.
	
	The propagation direction $\Ang$ shows a strong influence on both the mean bias and its variation. 
	This behaviour is likely due to spectral leakage, because spectral leakage appears  as ``streaks"  in the spectrum along the spectral axes $k_x$, $k_y$ and $\omega$ 
	(See illustration in figure \ref{fig:leakage_illustration2D}).
	The second column in figure \ref{fig:spec_ex} shows the extracted DSVs of tapered surface elevation data. The mean and random bias is strongly suppressed in all cases and the dependency on $\Ang$ is significantly reduced, compared to the analysis without a Hann window.
	Note that $\DAng=\ang{15}$ represents a rare, very narrow spectrum, but  the described effects are visible for all spectral widths, decreasing with increasing $\DAng$.
	
	We observe that for smaller wavenumbers ($k<1$), using a Hann window does not mitigate the effects of spectral leakage as strongly as for $k>1$. This is because the Hann window suppresses long-range spectral leakage (more than one-two pixels), while increasing the short-range spectra leakage  (The central lobe of the Hann window has a width of 2 bins instead of 1 for no-window). The ``steepness" $\del\omega/\del k$ grows with decreasing $k$, meaning that spectral intensity that leaks in the $k$-direction appears as a strong broadening in the $\omega$-direction at neighbouring $k$-values. In our simulations the spectrum has a peak at $\omega=1$ (thus close to $k=1$) and quickly decreases towards $k\approx 0.5$. Therefore, the combined effects of a steep slope in the energy spectrum and the dispersion relation cause the short-range spectral leakage to be most prominent for $k<1$.
	
	The effect of a steep spectral slope --- discussed further in section \ref{sec:peakedness} --- can be illustrated by considering that at, say,  $k=0.6$ the spectrum shows intensity at $\omega_0(0.6)$, but also intensity leaked from $\omega_0(0.6+\delta k)$ and $\omega_0(0.6-\delta k)$, the latter of which is much less than the former. The algorithm therefore finds a frequency between $\omega_0(0.6)$ and $\omega_0(0.6+\delta k)$, which gives a non-zero DSV (see also section \ref{sec:NSP_and_leakage}).

	\label{sec:DSV_Dang}
	\subsection{Influence of the mean propagation direction and angular spread}
	\label{sec:angwidth}
	
	To study the influence of the angular width on the mean and random bias in the extracted DSVs, simulations were run with three propagation directions $\Ang \in \{\ang{90},\ang{75},\ang{60}\}$ and a range of spectral widths $\DAng$, of which two representative examples, ($\ang{15},\ang{60}$), are shown in figure \ref{fig:spec_ex}  (see table \ref{tab:case_collection}, \caseName{1}, for all parameters).
	
	All cases show that an increase in spectral width $\DAng$ significantly decreases both mean and random biases across all wavenumbers. 
	Results for $\DAng > \ang{60}$ are not shown here, 
	as no notable further improvements were observed.
	
	The reduction in angular spread $\DAng$ leads to an increase in spurious DSVs $\vecDSV$, because the wave components of a narrow spectrum are mostly influenced by the current component parallel to the mean propagation direction $\Ang$. The narrower a spectrum is, the smaller an observable  change in frequency due to the perpendicular current component $\vecDSV_\perp$ becomes. 
	Therefore, the influence of false spectral intensity from any source (e.g. noise, aliasing, higher harmonics, spectral leakage) outside the real spectrum can lead to strong perpendicular DSV components when the spectrum is very narrow. For a more detailed explanation, see Fig.~\ref{fig:spec_ex}
 where
 we see the combined influence of a noisy spectrum  and spectral leakage.    
	
	Note, that all cases exhibit negligible spurious DSVs around $k\approx 1$. This is simply a consequence of the spectral peak lying near this value.
	
	The dependency of results on the mean propagation direction $\Ang$ was also investigated in a second manner: The videos were rotated numerically by angles $\Angg_\text{rot}$ up to $\ang{45}$ before repeating the DSV extraction. This is equivalent to rotating the camera in an experiment. Pairwise comparison of results with same values of $\Ang+\Angg_\text{rot}$ show no significant difference.
	
	Two conclusions can be drawn: First, 
 the only effect $\theta_\mathrm{rot}$ can have on physical quantities is a trivial rotation of the coordinate system, so any other $\theta_\mathrm{rot}$-dependencies
can be traced back to spectral leakage, which mainly occurs parallel to the
video's
axes, thus breaking rotational symmetry.
	Second, 
 will see in section \ref{sec:DSV_U} that when a current is present, the measurements can indeed differ after camera rotation, hence
 rotating the camera or data provides a useful check for spectral leakage.
	
	The use of a Hann window prior to DSV extraction greatly mitigates the effects of spectral leakage, as discussed before, pushing even the narrow spectrum cases down to sub-resolution (with respect to implied velocity resolutions $\dcdw$ and $\dcdk$).

	\subsection{Influence of peakedness \texorpdfstring{$\gamma$}{gamma}}\label{sec:peakedness}
	\label{sec:DSV_gamma}
	
	Depending on how developed a sea is, the best fit for the frequency spectrum uses a peak enhancement factor $\gamma$ between 1 and 7, where 3.3 is a commonly used value for most applications \cite{goda2010}. Recently, Mazzaretto et al. found that a global mean of $\gamma \approx 2.4$ is better suited \cite{MAZZARETTO2022112756}.
	
	Since the choice of $\gamma$  only affects the spectrum in a small range around the peak, it also offers itself as a tool to examine the influence of steep gradients in the spectrum.
	We therefore compare the DSVs for  $\gamma \in \{1...7\}$.  (see table \ref{tab:case_collection}, \caseName{2}, for all parameters).
	
\begin{figure*}[ht]
		\centering
		\includegraphics[width = \myFigWidthSingles]{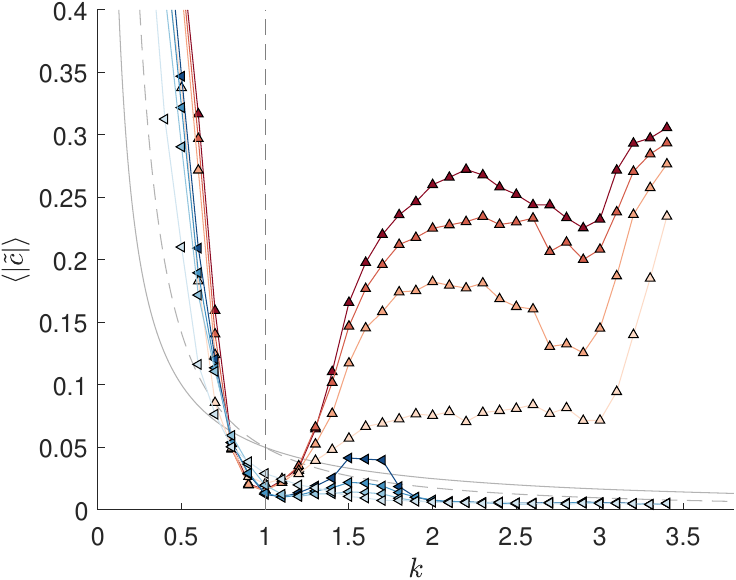}
		\includegraphics[width = \myFigWidthSingles]{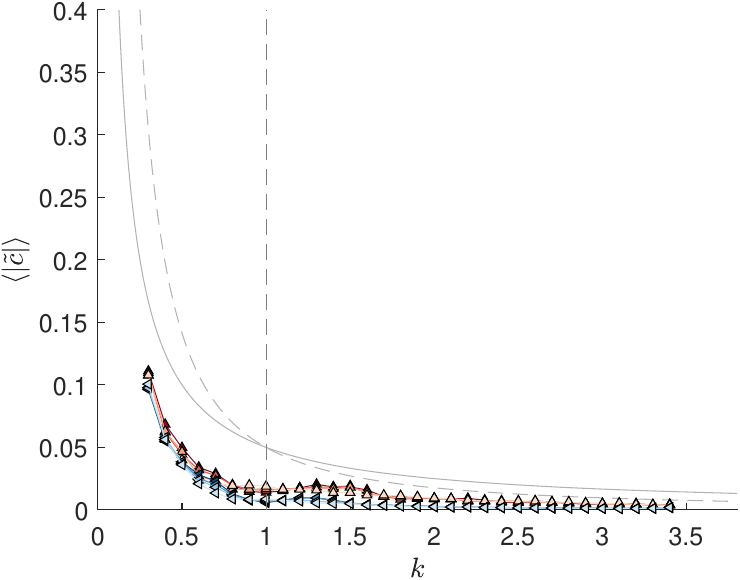} \\
		\includegraphics[width = \myFigWidthSingles]{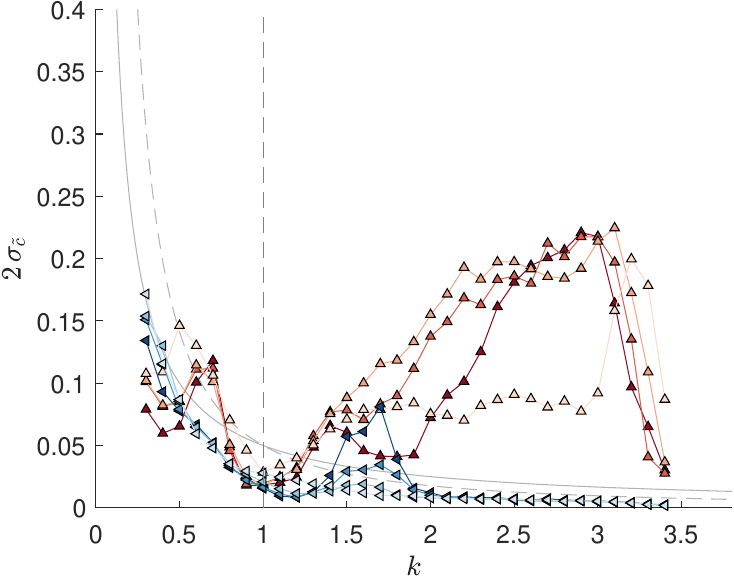}
		\includegraphics[width = \myFigWidthSingles]{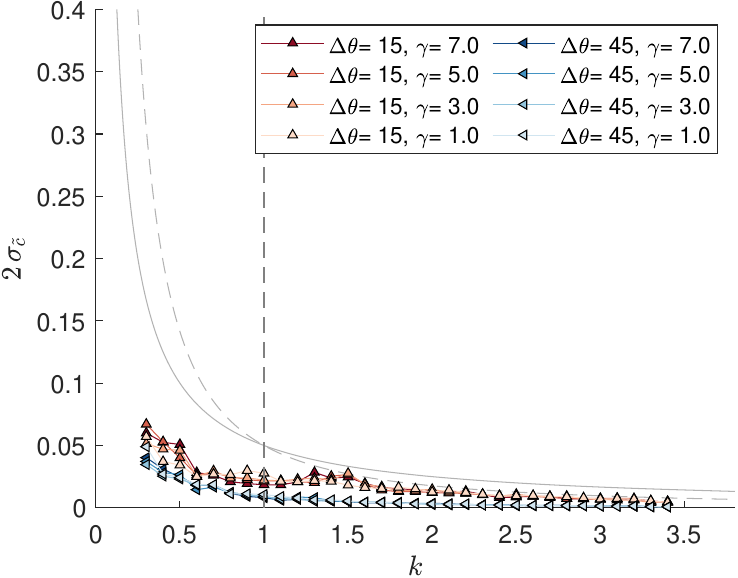}
		\caption{Same as figure \ref{fig:Ang_Dang} but for a selection of parameter combinations from the parameter set of\caseNameRobust{2} (see table \ref{tab:case_collection}).}
		\label{fig:peakedness}
	\end{figure*}

	The results presented in figure \ref{fig:peakedness} show that for spectra with small angular spread,  a more strongly peaked spectrum  (higher value for $\gamma$) causes increased biases (mean and random) across all wavenumbers except for a small range around the spectral peak at $k\approx 1$, where the relative increase in spectral intensity reduces the influence of spectral leakage from wavenumbers $k\neq 1$. 
	A wider angular spread of the spectrum, on the other hand,  mitigates this to a large degree, especially towards higher wavenumbers, where the influence of spectral leakage from the peak is nearly eliminated and sub-resolution biases are achieved. For angular spreads larger than $\DAng>\ang{45}$, the DSVs become independent of $\DAng$, though this also changes with spectral resolution.
	
	Towards lower wavenumbers ($k<1$), without a Hann window, any extracted DSV is unreliable, as spectral leakage dominates the result due to the exponential decrease of "real" spectral intensity towards $k=0.5$ and the steepness of the dispersion relation (see previous discussion in section \ref{sec:DSV_example}).
	
 Once again 
the use of a Hann window improves the results so far that random and mean biases are sub-resolution even for strongly peaked  ($\gamma=7$), narrow ($\DAng=\ang{15}$) spectra, effectively eliminating the dependence on $\DAng$, and to a degree, on resolutions (see section \ref{sec:size}).

	%

	\subsection{Influence of resolutions \texorpdfstring{$\delta k$, $\delta\omega$}{dk, dw}}\label{sec:resolution}\label{sec:duration}\label{sec:size}
	The range of spectral leakage is constant in terms of 
 the number of
 pixels or bins
 affected (see Fig.~\ref{fig:leakage_lobes})%
 , and thus the 
 leakage 
 range in $k$ (and $\omega$) is determined by the resolution $\delta k$ (and $\delta\omega$). 
	
	We therefore  vary the spatial size $L$ and temporal duration $T$ of the videos to change the frequency resolution $\delta \omega=1/T$ and wavenumber resolution $\delta k = 1/L$ to observe their influence on the DSV extraction 
	(see table \ref{tab:case_collection}, \caseName{3}, for all parameters).
	
\begin{figure*}[ht]
		\centering
		\includegraphics[width = \myFigWidthSingles]{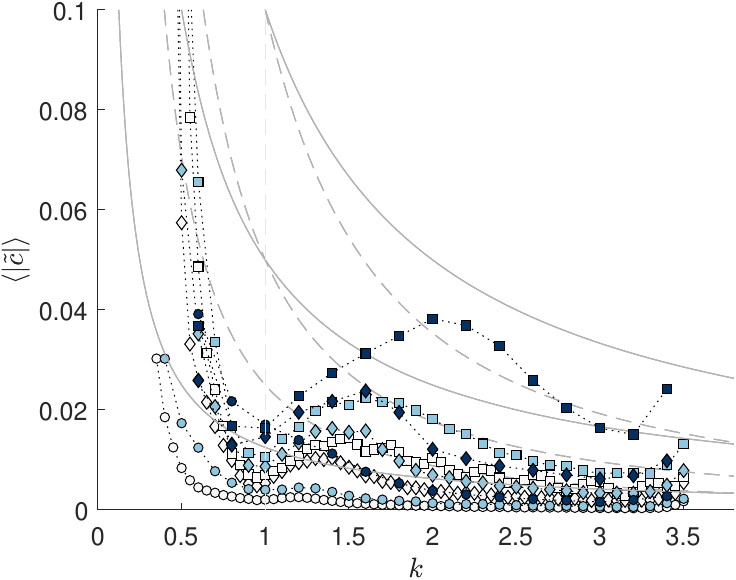}
		\includegraphics[width = \myFigWidthSingles]{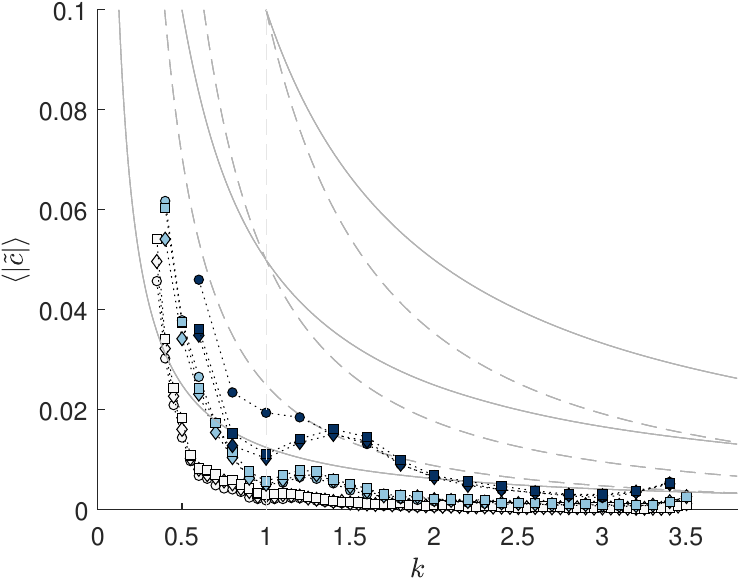} \\
		\includegraphics[width = \myFigWidthSingles]{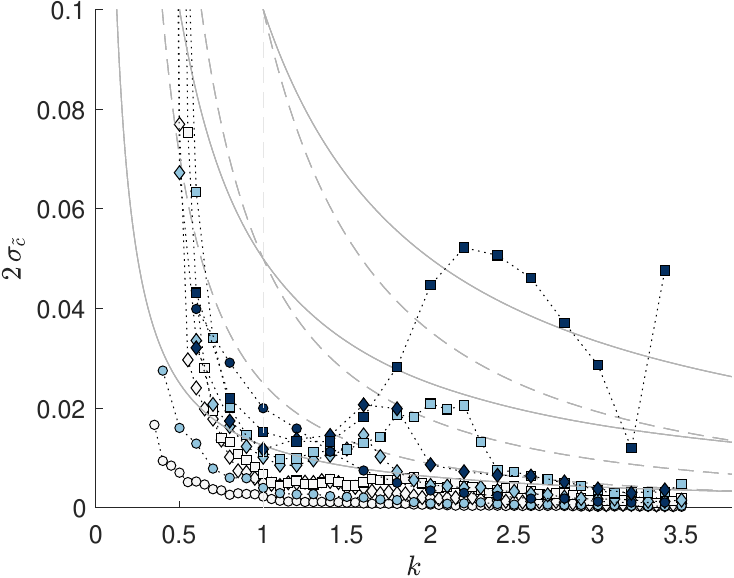}
		\includegraphics[width = \myFigWidthSingles]{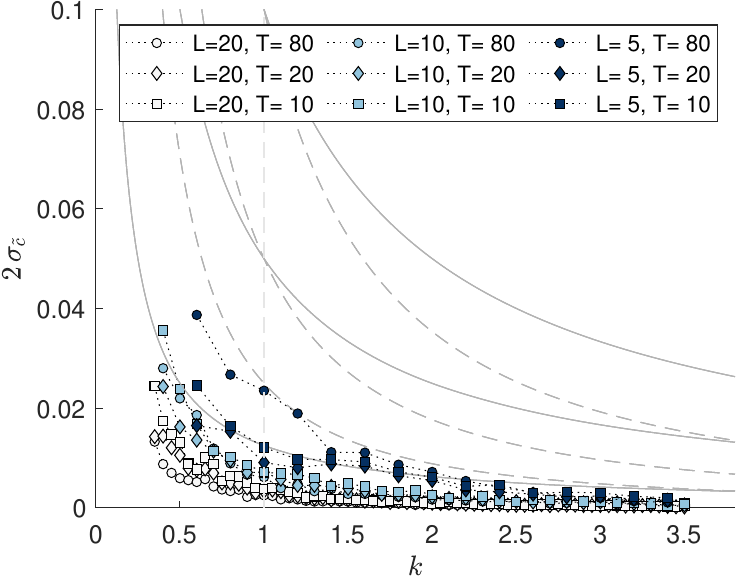}
		\caption{Same as figure \ref{fig:Ang_Dang} but with parameter set \caseNameRobust{3} (see table \ref{tab:case_collection}). As in figure \ref{fig:Ang_Dang} The solid and dashed lines correspond to the implied velocity resolutions from frequency and wavenumber resolution, respectively. From top to bottom the resolution increases, i.e. $\delta w$ or $\delta k$ become smaller. }
		\label{fig:video_duration}
	\end{figure*}

	For a spectrum with an angular spread of $\DAng=\ang{60}$ and a mean propagation direction of $\Ang=\ang{90}$ simulations were performed with a range of video lengths $T=\SIrange{5}{80}{}$ and domain sizes $L = \SIrange{5}{20}{}$. The resulting DSVs are shown in figure \ref{fig:video_duration}. (Note that the results for $T<10$ are not shown as they did not show relevant differences compared with results for $T=10$)
	
	Clearly, for untapered data, both improving  $\delta k$ or $\delta \omega$  reduces the biases in DSVs, albeit improving the wavenumber resolution $\delta k$ has a stronger influence.
	
	Note how the maximum in biases
 in Fig.~\ref{fig:video_duration}
 moves to higher $k$ with decreasing $T$. This is due to the characteristic function $G$ having a width $a$ proportional to frequency resolution, $a=4\delta\omega$. 
	When this width is reduced, the intensity that leaks from the spectral peak onto the cylinder at $k>1$ can fall outside the reach of the characteristic function, thus reducing or removing its influence.
	
	When the data is tapered using a Hann window, this effect is mostly eliminated, as the long-range spectral leakage is heavily suppressed. Moreover, the influence of frequency resolution $\delta \omega$  is strongly reduced (graphs of same color group together). 
	
	An exception to the rule of thumb that longer videos are always better can be seen for the longest videos $T=80$ on the smallest domain $L=5$. Here, the biases are actually greater than for the shorter cases with $T\leq 40$. This is due to the characteristic function
 $G$
 becoming narrow enough to not encompass the width of spectral intensity in the $\omega$-direction. In one dimension this would be proportional to $\delta\omega$, but in 2D or 3D can be dominated by leakage in $k$,  as this effective broadening in $\omega$ scales with $\mathcal{O}(\del\omega/\del k\ \delta k)= \mathcal{O}(\delta k/\sqrt{k})$. 
	
	Note, however, that the results presented 
 %
 in Fig.~\ref{fig:video_duration}
 all show biases well below the implied velocity resolutions, 
 provided
 the data is tapered using a Hann window.
	If one has to choose between increasing the domain size or the video duration, it is clear that an increase in domain size will give the most benefit.
	
	\subsection{Influence of a background current \texorpdfstring{$U_0$}{U} }\label{sec:current}
\label{sec:DSV_U}

The cases shown above are without a background current. The presence of a 
spatiotemporally
uniform current $\myVec{U}\neq 0$ breaks the rotational symmetry of the dispersion relation, because of the additional, angular dependent term $\veck\cdot\myVec{U}$
in equation \eqref{eq:SnJ}. Like all velocities, $\boldsymbol{U}$ is given in units of the phase velocity at the spectral peak (see Section \ref{sec:dimensions})%
. 

To see the effect of a current on the biases in the extracted DSVs, we assume a (vertically and horizontally) constant current $\myVec{U}$ with directions $\Angg_U=\ang{90},\ang{0},\ang{-90}$ and current strengths in the range $|\myVec{U}|=\SIrange{0.025}{0.15}{}$ (see table \ref{tab:case_collection}, \caseName{5}, 
for all parameters).
The input wavenumber spectrum has an angular spread of $\DAng = \ang{60}$, and a peakedness of $\gamma=3.3$, representing a realistic scenario. The mean propagation direction is held at $\Ang=\ang{90}$, resulting in following, crossing and opposing current, respectively.

\begin{figure*}[ht]
	\centering
	\includegraphics[width = \myFigWidthSingles]{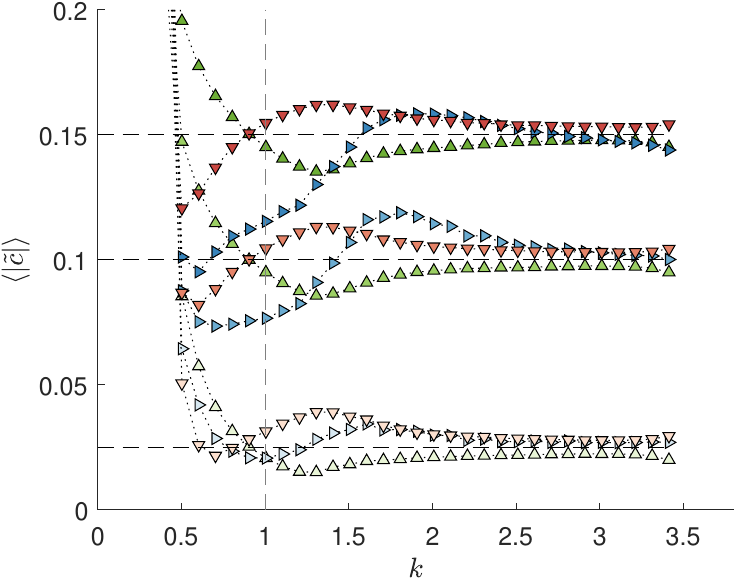}
	\includegraphics[width = \myFigWidthSingles]{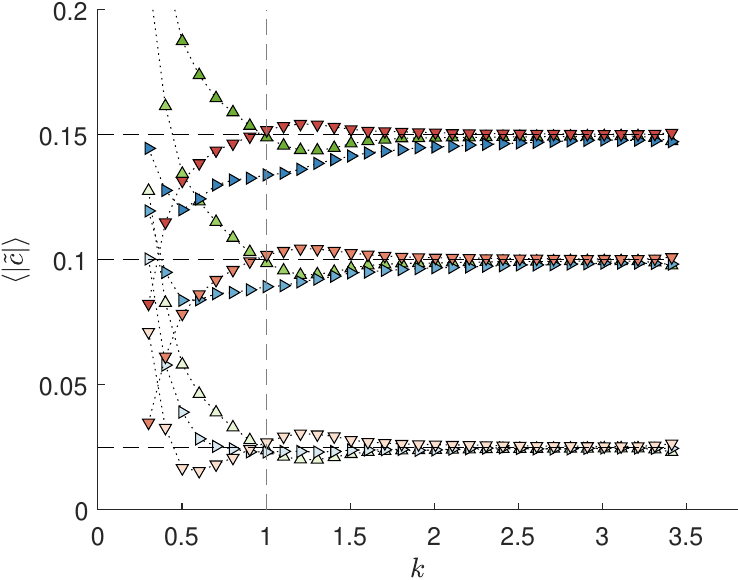} \\
	\includegraphics[width = \myFigWidthSingles]{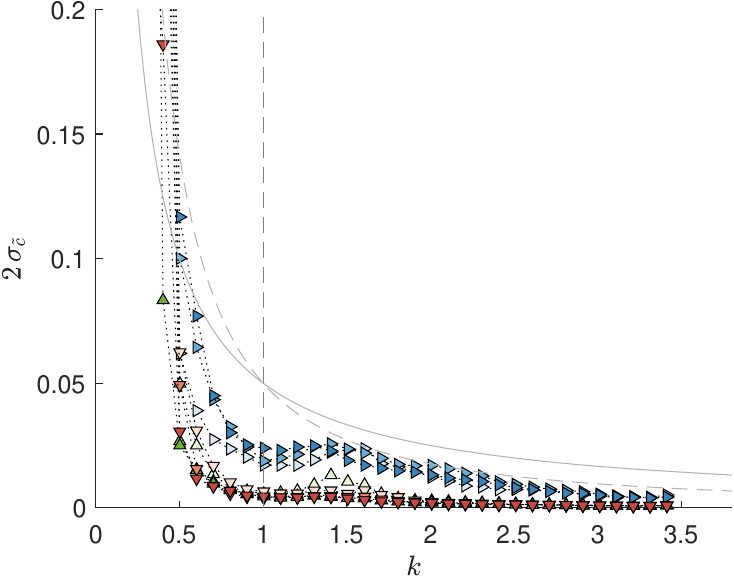}
	\includegraphics[width = \myFigWidthSingles]{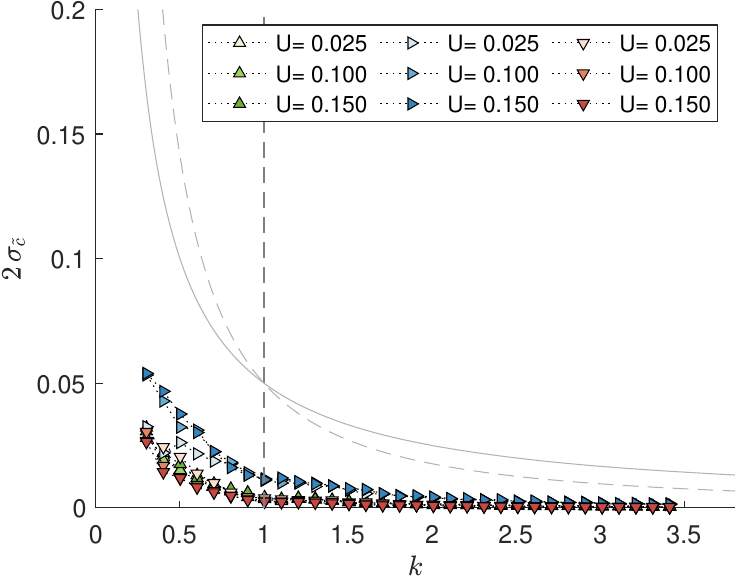}
	\includegraphics[width = \myFigWidthSingles]{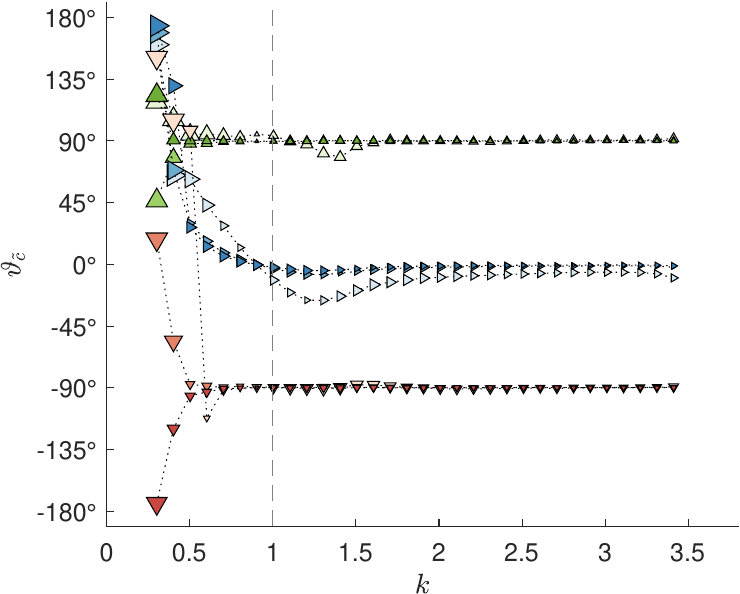}
	\includegraphics[width = \myFigWidthSingles]{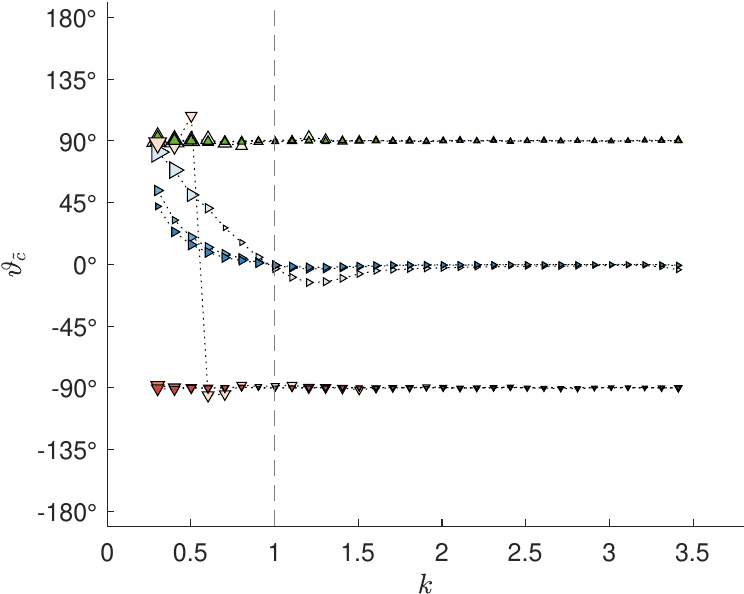}
	\caption{Same as figure \ref{fig:Ang_Dang} but with parameter set \caseNameRobust{5} (see table \ref{tab:case_collection}). The dashed horizontal lines represent the three background velocity values. Deviations from these are spurious. The third row shows the direction $\myvartheta_{\vecDSV}$ of the average Doppler shift velocity.}
	\label{fig:ConstCurr}
\end{figure*}

The extracted DSVs are shown in figure \ref{fig:ConstCurr}.
For $k>1$, we see a bias toward wave-opposing DSVs, while for $k<1$, we see a bias toward wave-following DSVs.
This can be seen from over-/underestimations for $\myVec{U}$ antiparallel/parallel to the wave propagation direction, respectively. The absolute magnitude of errors is similar to the case of quiescent water (see section \ref{sec:resolution}).
In the case of a pure cross-current, however, this also results in a turning of the DSVs toward $\myvartheta_{\vecDSV}=\pm \ang{90}$ for wavenumbers around the spectral peak $k\lesssim 1.5$, as can be seen in  the third row of 
panels
in figure $\ref{fig:ConstCurr}$. Particularly, where the spectral intensity falls off to zero rapidly ($k\approx 0.5$), leakage from the peak of the spectrum can dominate the signal, biasing the direction of DSVs towards the
wave
propagation direction, i.e. $\ang{90}$. 

Tapering the data using a Hann window mitigates these biases in amplitude and direction as in the case of quiescent water. For a cross-current, however, the mean bias remains significant, i.e. on the order of 10\% of the phase velocity at the spectral peak and unusually also in the vicinity of the spectral peak.

The error in the direction $\myvartheta_{\vecDSV}$ of DSVs  for wavenumbers below the peak is also reduced significantly for  following and opposing currents. In the case of a cross-current these errors remain significant, (\ang{10}-\ang{45}) if the current is relatively weak or the wavenumber goes toward $k=0.5$. 

\begin{figure*}[ht]
	\centering
	\includegraphics[width = \myFigWidthSingles]{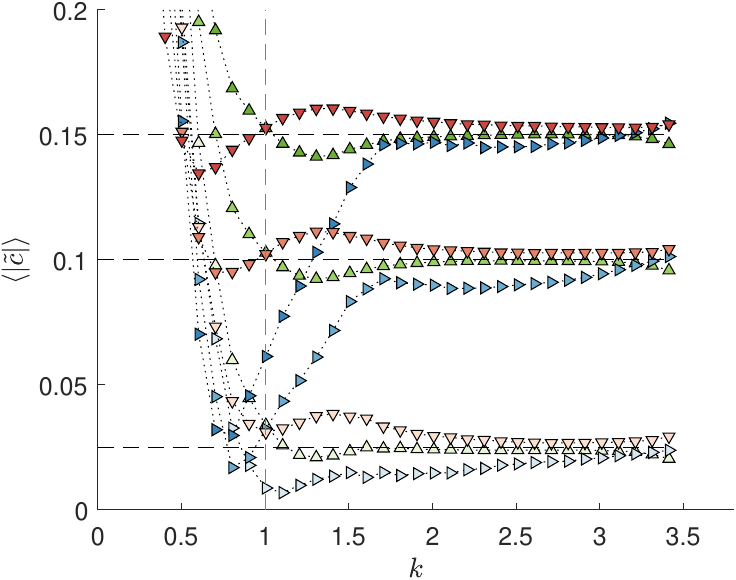}
	\includegraphics[width = \myFigWidthSingles]{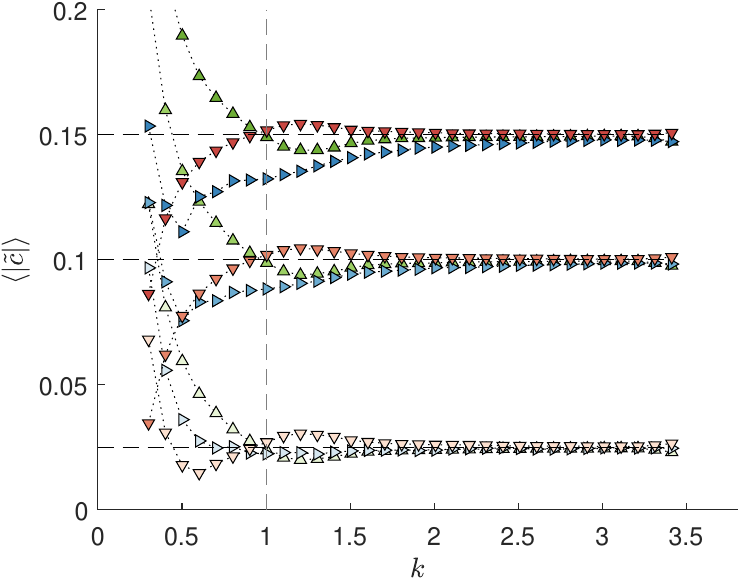} \\
	\includegraphics[width = \myFigWidthSingles]{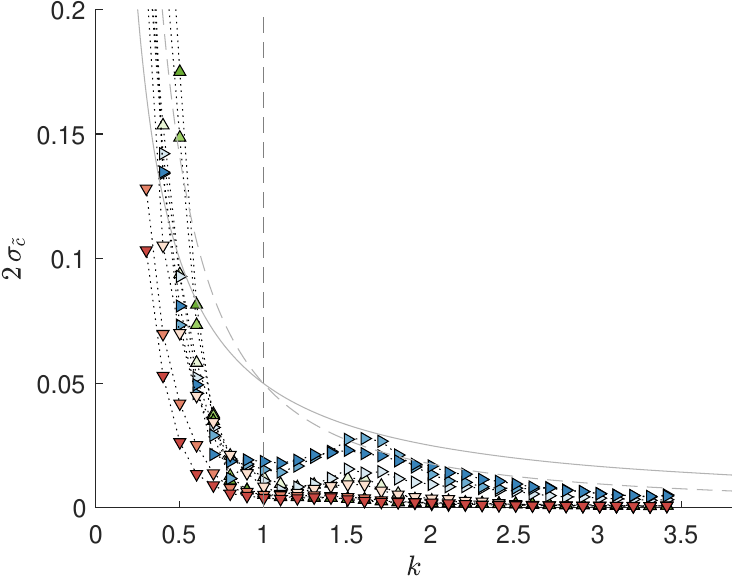}
	\includegraphics[width = \myFigWidthSingles]{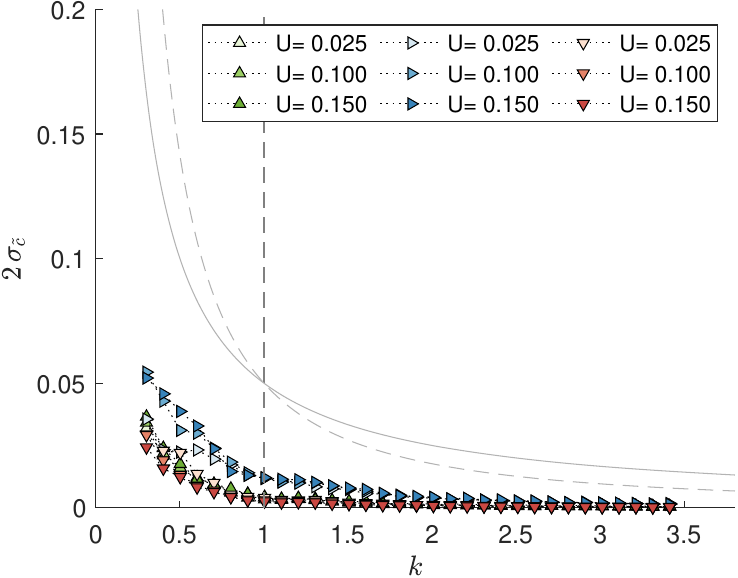}
	\includegraphics[width = \myFigWidthSingles]{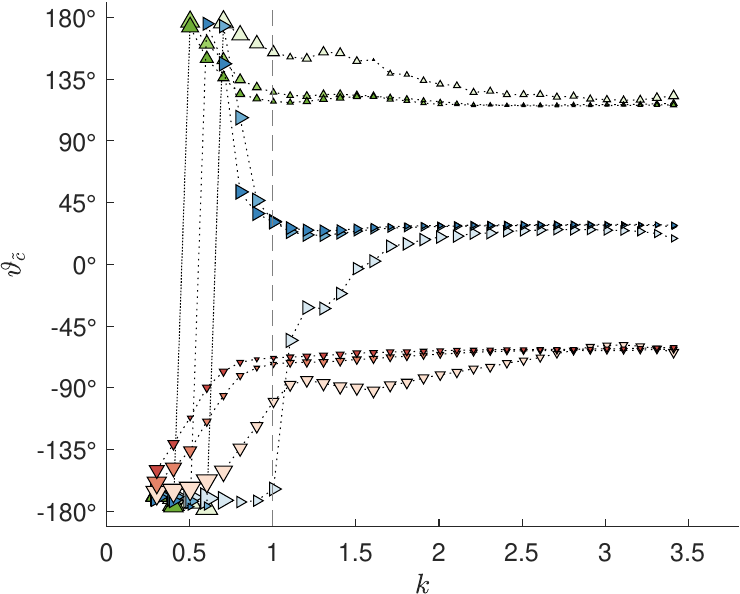}
	\includegraphics[width = \myFigWidthSingles]{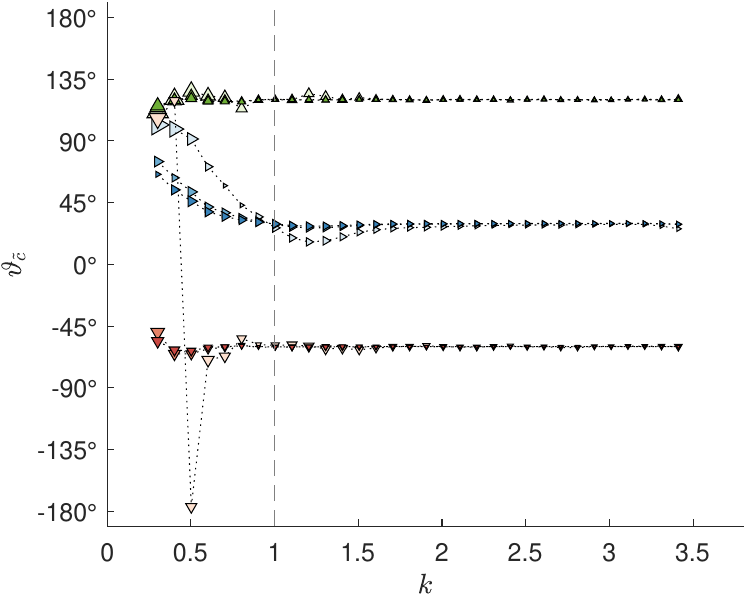}
	\caption{Same as figure \ref{fig:ConstCurr} but with a \ang{30} rotated video. Note, how the case of cross currents (blue right facing triangles) result in unusable DSVs $\vecDSV$ especially around the spectral peak, when the data is untapered. (top left) }
	\label{fig:ConstCurr_rot30}
\end{figure*}

As mentioned in section \ref{sec:angwidth}, turning the camera in an aerial measurement can reveal the influence of spectral leakage (one could also rotate the resulting images, bearing in mind that this is not in general a lossless operation and might also reduce the field of view).
Figure \ref{fig:ConstCurr_rot30} shows the extracted DSVs for a case with the same parameters as in figure \ref{fig:ConstCurr}, but with current and spectrum rotated by \ang{30} (Similar results were obtained for \SIrange{15}{45}{\degree}).

For the untapered data, we find even stronger mean biases after rotation, especially for a cross-current. 
Here the DSVs around the spectral peak at $k\approx 1$ drop below \SI{50}{\percent} of the background current, usually an unacceptable level of error. However, the use of a Hann window again mitigates this, to the point where the results of different rotation angles become virtually indistinguishable.

\section{Conclusions}\label{sec:conclusions}

We have investigated biases from spectral leakage in remote sensing of currents from analysis of wave spectra. Apparent, spurious Doppler shifts in the phase velocity are observed even in the absence of a current, and we analyse how these depend on wave spectrum and the properties of the signal processing procedure. Synthetically generated surface elevation data were used to simulate a random sea state adhering to a JONSWAP wave spectrum \cite{Hasselmann73} with a $\cos^2$ directional distribution, resulting in cubes of data (videos). These were subsequently analysed with different methods in common use to extract the Doppler-shift velocities (DSV). Following reference \cite{Smeltzer19}, an appropriate measure of random errors and biases are the ``Doppler shift resolutions'' corresponding to the change in inferred velocity due to a shift of one pixel in wavenumber or frequency. 

A comparison of the normalized scalar product approach (e.g.\ \cite{Smeltzer19}) and a least-squares method for extracting DSV showed that the former is preferable in all cases, and was therefore used for all subsequent analysis herein. 

Assuming the simplest case of quiescent water (i.e.\ no current), a complex interplay is found between wave-spectrum width and peakedness, and the wave-vector and frequency resolution, together affecting the nature and extent of spurious Doppler-shift ``measurements''. 
Spectral leakage causes greater problems when the wave spectrum is strongly peaked and highly directional so that areas of the observed frequency-wave vector spectrum which are important to Doppler-shift extraction have very low signal. 
Conversely, when the angular spread is wide, $\DAng>\ang{60}$ ($\DAng$: full width of angular distribution
), spurious DSVs are small, i.e., sub-resolution. For strongly directional spectra  $\DAng<\ang{60}$,  severe biases emerge, with amplitudes on the order of the group velocity, depending on the JONSWAP peakedness parameter $\gamma$. 
The biases are sensitive to resolution in frequency and wave-number space, and especially in the absence of tapering (see below) poorer resolution rapidly leads to unusable data for narrow and strongly directional wave fields. Biases are most severe at low wavenumbers compared to 
that at
the spectral peak.

Tapering the video cubes with a 3D Hann window (e.g.\ \cite{Nuttall1981}) lowers the biases to the velocity resolution level implied by the wavevector and frequency resolutions --- $\dcdk$ and $\dcdw$ as defined in equation \eqref{eq:dcdwk} --- or even below. Indeed, the effect of $\omega$-resolution is mostly removed for tapered data (note that although biases are now sub-resolution, the resolution itself will eventually be too poor for purpose). This implies that in data acquisition one should prioritize large areas rather than longer time series if spectral leakage is a problem. 

The effect of spectral leakage is most pronounced in the $k_x$ and $k_y$ directions in the spectrum, and hence depends on the angle of propagation relative to these. The biases increase towards an angle of $\Ang=\ang{45}$  for all wavenumbers outside a small range around the spectral peak $k=k_p$. This dependence on $\Ang$ is also mostly removed by tapering the data. Clearly the observed current velocity cannot depend on which way the camera is held, meaning that comparison with results when the video is rotated%
, say,
$\ang{45}$ could give a simple indication of the severity of spectral leakage problems. 

For the case of a constant background current $\myVec{U}$ with strengths up to $0.15\, c_0(k_p)$ a strong dependence on the angle between the current and the waves is observed.  While the magnitudes of the spurious DSVs are mostly 
smaller than
the implied velocity resolutions, we find a significant bias in the direction of the DSVs around $k=k_p$ when the waves propagate perpendicular to the current. This is also mitigated by tapering the data, but not removed in the case of a cross current.

\subsection{Recommendations for mitigation}

Summarising the outcome of our analysis from a practical viewpoint we offer the following considerations to mitigate the errors and biases related to spectral leakage in remote sensing of currents from observed wave spectra.
In the extraction of Doppler shift velocities from the spectrum, the commonly used least-squares method is not recommended except if calculation cost is a severe restriction; a normalized scalar product procedure gives universally better results (other methods are also in use, but were not tested).
Tapering the spatio-temporal data with a  3D Hann window greatly reduces the mean and random biases and 
their dependence
on spectral shape, spectral resolution, and camera orientation. When errors due to spectral leakage are suspected,
rotating the camera (either the actual camera or the resulting images) by 
some intermediate
angle before analysis (we used \ang{30}) and comparing results could reveal whether long range spectral leakage is causing spurious results, because leakage mainly occurs along the axes of the images. We find that the wavenumber resolution plays a more important role in the DSV biases than frequency resolution. The influence of the latter can be nearly eliminated by tapering the data. Increasing the spatial domain size to improve wavenumber resolution yields the largest improvement and should therefore be prioritized over longer time series if spectral leakage is a concern.
The Doppler shift resolutions $\dcdw$ and $\dcdk$ defined in equation \eqref{eq:dcdwk} are useful as conservative measures of errors and biases due to limited resolution and spectral leakage.

\section{Acknowledgments}
\noindent The authors would like to thank Drs. Susanne St\o le-Hentschel, Luc Lenain and Nick Pizzo for 
discussions.

\renewcommand\thesection{\Alph{section}}
\setcounter{section}{0}

\appendix

\subsection{Comparison of window functions}\label{app:window_comparison}
\begin{figure}[h]
	\centering
	\includegraphics[width = \myFigWidthSingles]{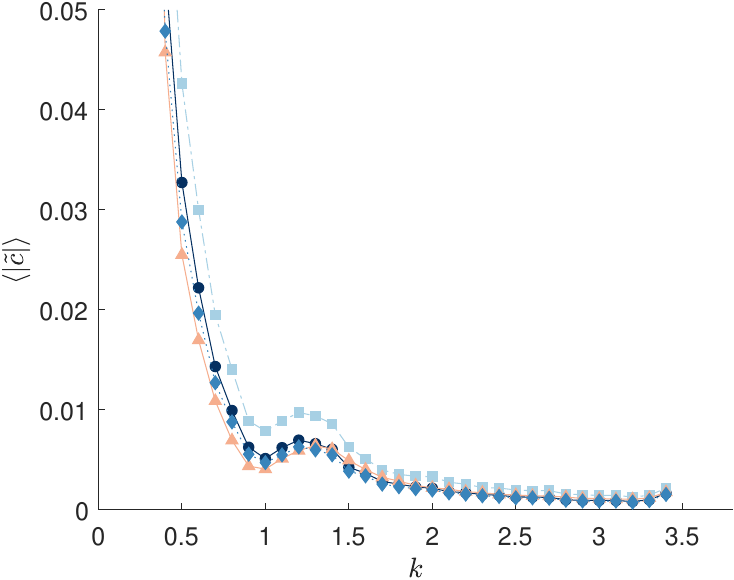}\\
	\includegraphics[width = \myFigWidthSingles]{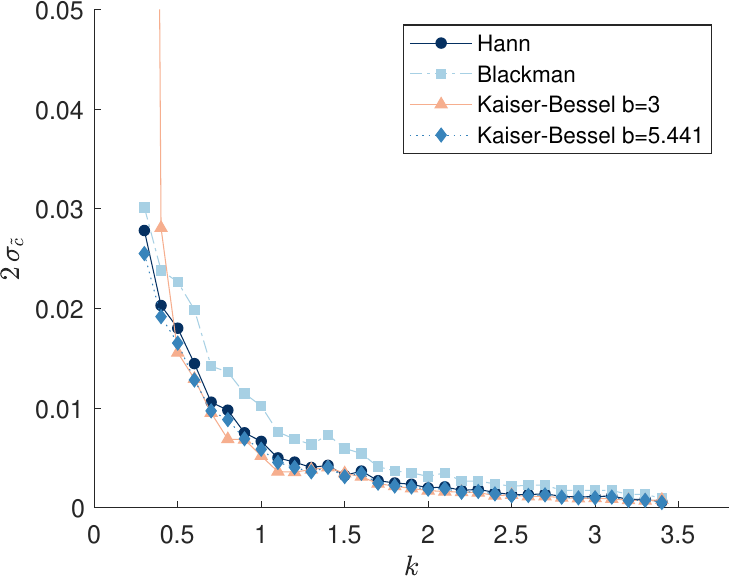}
	\caption{%
		Doppler shift velocities (DSV) 
  in units of the peak phase velocity
  for a normal parameter combination ($\Ang=\ang{90}$, $\DAng=\ang{60}$, $\gamma=3.3$, $L=10$, $T=20$, $U=0$) obtained using the NSP method after tapering the data with different window functions. The top and bottom graph show the DSVs $\vecDSV$  in terms of the mean (top) and standard deviation (bottom), respectively. 
	}
	\label{fig:DSV_win_comp_results}
\end{figure}

There exists a wide variety of window functions employed 
across disciplines,
as discussed in e.g. \cite{Nuttall1981}. 
The suppression of the long range spectral leakage is the most important beneficial effect of using a window function, but stronger suppression comes at the cost of increasing the short range leakage. For example, the Blackman window has a central lobe that is about 20\% wider than that of the Hann window, as can be seen in figure \ref{fig:leakage_lobes}. To examine its detrimental effect, the extraction of DSVs has been repeated with a selection of commonly used window functions. As the results are very similar we only show the results for the Hann window
\begin{equation}
	w_{\text{}}(t)  = 0.5- 0.5\,\cos\left(2\pi\,t/T\right)  ,
\end{equation}
the approximate Blackman window,
\begin{equation}
	w_{\text{}}(t)  = 0.42- 0.5\,\cos\left(2\pi\,t/T\right) +0.08\,\cos\left(4\pi\,t/T\right)  ,
\end{equation}
and the Kaiser-Bessel window
\begin{equation}
	w_{\text{}}(t)  = \frac{1}{T}I_0\left(b\sqrt{1-(2\pi\,t/T)^2}\right)  ,
\end{equation}
with $I_0$ being the zeroth order modified Bessel function of the first kind and $b$ a free parameter controlling the width of the central lobe. In figure \ref{fig:DSV_win_comp_results} the extraced DSVs in terms of their main and random bias are shown. The differences in the results are insignificant, except for the Blackman window showing systematically higher biases due to its wider central lobe. 

\subsection{Details of numerical implementation}\label{app:numdetails}

Here follow further details on the implementation of the NSP method in section \ref{sec:NSP}. 

\subsubsection{NSP, LS and discretized data}
Because of discretization, obtaining the  spectral intensity $P_i$ 
on a cylinder surface needs to be replaced with the spectral intensity in a volume around it, i.e. a cylinder shell, containing a wavenumber bin around $k_i$:
\begin{equation}
	F_i(\mathbf{k}, \omega) = 
	\begin{cases}
		\sqrt{P(\mathbf{k}, \omega)}, & \text{if } |k-k_i|\leq \tilde{\Delta} k\\
		0,   & \text{otherwise}
	\end{cases}.
	\label{eq:g}
\end{equation}
This also renders the characteristic function $G_i$ (equation \eqref{eq:DSVkernel}) a function of $k$. The integrals $\myMean{...}$ in equation \ref{eq:normSp} then imply an additional integration over $k$ within the bin $k_i\pm\tilde{\Delta}k$.

The concept of the NSP method stays the same, with two details added: 
First, since the optimization parameter pair $\vecDSV$ is assumed constant within a wavenumber bin, the extracted DSV is a weighted average within that bin, effectively smoothing the function $\vecDSV(k)$. This also holds for the LS method, but does not change equation \eqref{eq:LS1}, as it only increases the number of triplets $(k_{x,j},k_{y,j},\omega_j)$.
Second, a new free parameter is introduced with the bin width $2\tilde{\Delta}k$ that needs to be chosen carefully. 
In this work, we use $\tilde{\Delta}k=2\delta k$, with $\delta k=1/L$ the wavenumber resolution, which is a compromise between increased smoothing (too large $\tilde{\Delta}k$) and strong noise, occurring when too few pixels of the spectrum lie within a bin. (One could employ an interpolation scheme to circumvent this; the algorithm used herein simply masks the data, see equation $\ref{eq:g}$)

Since the data are effectively averaged over $k$ with a running average of width $2\tilde{\Delta}k$, the DSVs were extracted such that two consecutive shells have an overlap of $3/4$, i.e.  $k_{i+1}-k_i = \delta k$.

\subsubsection{Width of the characteristic function}

As mentioned in section \ref{sec:parameters} the characteristic function $G$ (equation \ref{eq:DSVkernel}) used 
in fitting the dispersion relation to the measured spectrum
contains one free parameter%
, $a$,%
that determines its width. The choice of this parameter is somewhat delicate, as too small a value causes single intensity pixels (often outliers) or small high intensity regions to dominate the determination of the best fit. This is especially problematic for small ($k<1$) wavenumbers, when  the spectral leakage from  a noise-enhanced pixel near the spectral peak at $k\approx 1$ causes a small high-intensity region on the cylinder shell, whereas the ``real" spectral intensity, not originating from spectral leakage, is strongly broadened, thus having a larger total intensity but a smaller maximum intensity. In this case, the algorithm effectively ignores the real spectral intensity and leads to huge biases. On the other hand, too large a value for $a$ also leads to an increased influence of spectral leakage, as a too wide Gaussian is insensitive to shifts in its position. A shift with no, or only small penalty to the overlap with the real spectral intensity, that increases the overlap with intensity from leakage, is therefore more likely, also leading to biases.

To decide on a good compromise 
we fit the characteristic function to the Fourier transform of a Hann window, with $a$ as a free parameter. The result is $a\approx 1.62\,\delta\omega$. To 
ensure the width is not too small 
this value is roughly doubled to $a=4\,\delta\omega$. (Values for $a/\delta\omega$ between 2 and 6 were tried as well, but the results were most stable  between 3 and 5.)

Note, that if spectral leakage occurred only in the $\omega$ direction, $a=\delta\omega$ would be the ideal choice. However, as this is not the case, broadening in the $\omega$ direction originates not only from leakage in the $\omega$ direction, but also from leakage in the $k$-direction. The steeper the dispersion relation $\omega_\text{DR}
(k)
$, the more this effective $\omega$-leakage from $k$-leakage increases. A possible improvement of the used NSP algorithm would therefore be to use an adaptive $a$, that increases with $|\del\omega/\del k|$. In the interest of limiting the parameter space, it was deemed necessary to stick to a single value for $a$.

\subsection{Comparison NSP and a basic LS algorithm}
\label{sec:LSvsNSP}
Simple least squares based methods have by now been discarded in most applications. However, their low computational cost and ease of implementation are benefits to consider. We therefore repeated the DSV extraction using LS and compare the results and performance.
The least squares method used, like the NSP method described in section \ref{sec:doppler}, first singles out the spectral intensity on a cylinder surface with radius $k$. We discard data points with an intensity $P<0.2$ (after normalization) and obtain a list of $(k_{x,j},k_{y,j},\omega_j)$. On these, the cost function
\begin{equation}\label{eq:LS1}
	C(\vecDSV)=\sum_j (\omega_{DR}(\veck_j;\vecDSV)-\omega_j)^2
\end{equation}
is minimized to obtain the DSV $\vecDSV$. Repeating this for a list of wavenumbers $k$ yields 
the desired value of $\vecDSV$ for each value of $k$.
As in the NSP method and, the minimization/optimization step is performed using the Nelder-Mead simplex method \cite{Lagarias1998}.

\begin{figure}[ht]
	\centering
	\includegraphics[width = \myFigWidthSingles]{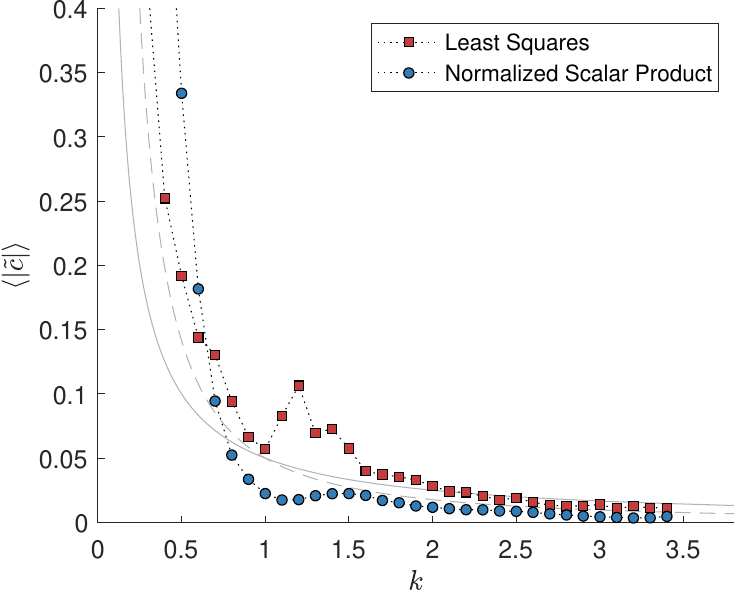}\\
	\includegraphics[width = \myFigWidthSingles]{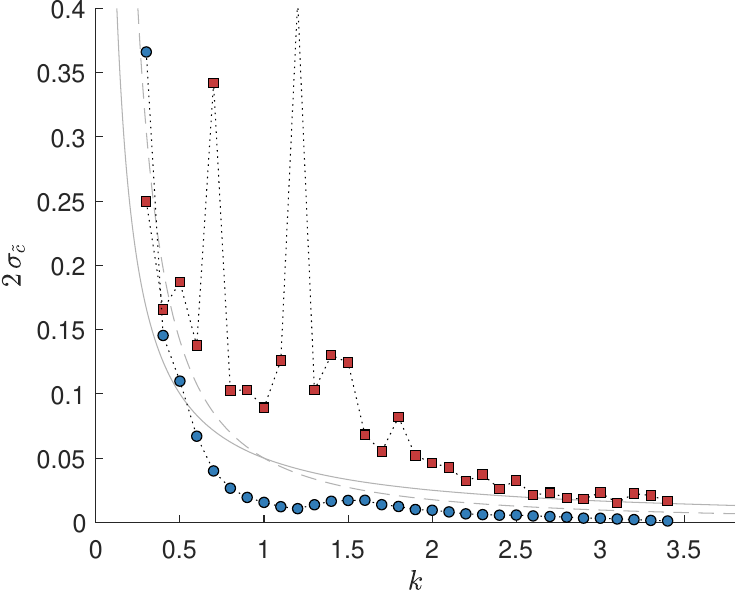}
	\caption{%
		Doppler shift velocities (DSV) 
  in units of peak phase velocity 
  for the parameter set  \caseNameRobust{4} (see table \ref{tab:case_collection}) extracted using the NSP and a least square method. The top and bottom graph show the DSVs $\vecDSV$ for quiescent water in terms of the mean (top) and standard deviation (bottom). The solid and dashed line correspond to the velocity resolutions implied by  frequency and wavenumber resolution, respectively.%
	       }
	\label{fig:ex_NSP_vs_LS}
\end{figure}

We find that the LS method runs significantly faster (up to a factor of $3$); clearly the difference in cost will depend on the hardware used as well as the implementation.
In terms of accuracy and precision, we find that the LS method performed consistently worse, and never better, than the NSP method
for practical purposes%
.  Figure \ref{fig:ex_NSP_vs_LS} shows an example result for a test case in quiescent water (see table \ref{tab:case_collection}, \caseName{4}, for all parameters). As can be seen, the DSVs obtained via the LS method can show both a mean  and random bias that exceed the implied velocity resolutions $\dcdk$ and $\dcdw$ for most wavenumbers, while the NSP method delivers sub-resolution DSVs for all $k>1$. It is worth pointing out, that for wavenumbers $k<1$ the NSP performs worse than the LS method. However, the random bias in that range is so large as to make both methods unusable.
Note that a more advanced LS method may perform with similar accuracy and precision, as is indicated in e.g.\ \cite{Huang12}. This would, presumably, lead to a computational cost similar to that of the NSP method, eliminating the advantage.
An iterative LS method has been compared with NSP by Huang et al. (2012). For waves on currents without vertical shear the methods performed similarly.

\section*{Data availability}
The data that support the findings of this study are available from the authors upon reasonable request.

\bibliographystyle{IEEEtran}
\bibliography{ref.bib}

\end{document}